\documentclass[twocolumn,trackchanges,times,tighten]{aastex631}

\usepackage[utf8]{inputenc}
\usepackage{amsmath}
\usepackage{multirow} 
\usepackage[ruled,vlined]{algorithm2e}
\usepackage{array}
\usepackage{color}

\usepackage{tablefootnote}

\usepackage{mathrsfs}
\usepackage{xcolor}
\definecolor{light-gray}{gray}{0.95}
\newcommand{\code}[1]{\colorbox{light-gray}{\texttt{#1}}}

\usepackage{textcomp}
\usepackage{gensymb}
\usepackage{graphicx}

\received{}
\revised{}
\accepted{}

\shorttitle{Pulscan: Binary pulsar detection using unmatched filters on NVIDIA GPUs}
\shortauthors{White J., Ad\'amek K., Roy J., Ransom S., Armour W.}

\begin{document}
\title{Pulscan: Binary pulsar detection using unmatched filters on NVIDIA GPUs}

\correspondingauthor{Wesley Armour}
\email{wes.armour@oerc.ox.ac.uk}

\author[0000-0003-2690-6858]{Jack White}
\affiliation{Oxford e-Research Centre, Department of Engineering Science, University of Oxford, 7 Keble Road, Oxford OX1 3QG, United Kingdom}

\author[0000-0003-2797-0595]{Karel Ad\'{a}mek}
\affiliation{Oxford e-Research Centre, Department of Engineering Science, University of Oxford, 7 Keble Road, Oxford OX1 3QG, United Kingdom}
%\affiliation{Department of Physics, Silesian University in Opava, Bezru\v{c}ovo n\'{a}m., 746 01, Opava, Czech Republic}

\author[0000-0002-2892-8025]{Jayanta Roy}
\affiliation{National Centre for Radio Astrophysics (NCRA-TIFR), Pune 411 007, India}

\author[0000-0001-5799-9714]{Scott M. Ransom}
\affiliation{National Radio Astronomy Observatory, Charlottesville, VA 22903, United States }

\author[0000-0003-1756-3064]{Wesley Armour}
\affiliation{Oxford e-Research Centre, Department of Engineering Science, University of Oxford, 7 Keble Road, Oxford OX1 3QG, United Kingdom}

\begin{abstract}

The Fourier Domain Acceleration Search (FDAS) and Fourier Domain Jerk Search (FDJS) are proven matched filtering techniques for detecting binary pulsar signatures in time-domain radio astronomy datasets. Next generation radio telescopes such as the SPOTLIGHT project at the GMRT produce data at rates that mandate real-time processing, as storage of the entire captured dataset for subsequent offline processing is infeasible. The computational demands of FDAS and FDJS make them challenging to implement in real-time detection pipelines, requiring costly high performance computing facilities. To address this we propose Pulscan, an unmatched filtering approach which achieves order-of-magnitude improvements in runtime performance compared to FDAS whilst being able to detect both accelerated and some jerked binary pulsars. We profile the sensitivity of Pulscan using a distribution (N = 10,955) of synthetic binary pulsars and compare its performance with FDAS and FDJS. Our implementation of Pulscan includes an OpenMP version for multicore CPU acceleration, a version for heterogeneous CPU/GPU environments such as NVIDIA Grace Hopper, and a fully optimized NVIDIA GPU implementation for integration into an AstroAccelerate pipeline, which will be deployed in the SPOTLIGHT project at the GMRT. Our results demonstrate that unmatched filtering in Pulscan can serve as an efficient data reduction step, prioritizing datasets for further analysis and focusing human and subsequent computational resources on likely binary pulsar signatures.

\end{abstract}

\section{Introduction}

Detecting pulsars in binary systems is crucial because they serve as unique laboratories for testing theories of gravity. Their presence in extreme gravitational environments, combined with their function as extremely precise celestial clocks, allows for highly accurate timing measurements of their dynamics.

Detecting solitary pulsars is possible by searching the frequency spectrum of a time domain signal captured by a radio telescope for peaks corresponding to the fundamental and higher harmonic frequencies of the spin frequency of the pulsar. However, in binary pulsar systems the relative motion between the pulsar and the observer, due to the orbit of the pulsar around its companion, leads to a Doppler shift in the observed frequency of the pulsar's signal. The Doppler shift spreads the frequency response of the pulsar across a range of frequencies, making detection more complex. It is more complex due to the reduced local prominence caused by having the same total power spread across multiple frequency bins, in contrast to all the power being in a single frequency bin.

The Fourier Domain Acceleration Search (FDAS) was proposed in \cite{Ransom2001}. It is based on matched filtering which operates on the frequency spectrum of the observed radio signature and uses the Fast Fourier Transform (FFT) to accelerate the convolutions required for the matched filtering process.

Whilst the FFT is an $O(N log N)$ algorithm for calculating the Discrete Fourier Transform (DFT), and so provides favourable performance compared to directly calculating the Discrete Fourier Transform using an $O( N^2 )$ algorithm, the key computational challenge lies in the convolution process with typical searches including over 100 matched filters.

In this paper we present an investigation of replacing the matched filters in FDAS with boxcar filters, and whether this new approach retains the ability to detect binary pulsars with the significant advantage of reduced execution time, allowing greater volumes of data to be processed in real time. We present multiple implementations of the algorithm, optimised for both CPU and GPU hardware. The implementations of Pulscan discussed in this paper are available in our freely accessible repository.\footnote{Repository: \href{https://github.com/jack-white1/pulscan}{https://github.com/jack-white1/pulscan}}

\subsection{SPOTLIGHT project at the GMRT}
\label{NSM-GMRT}

SPOTLIGHT\footnote{Website: \href{https://spotlight.ncra.tifr.res.in/}{https://spotlight.ncra.tifr.res.in/}}is a commensal survey instrument at the GMRT (funded by the National Supercomputing Mission, Govt. of India) aiming to discover a large population of FRBs associated with its host galaxies and pulsars. While the GPU-based petaflop-scale HPC facility is primarily designed for a real-time FRB search over $\sim$2000 beams, the facility will also be equipped with capacity to perform a matched filtering based pulsar search using FDAS/FDJS over 10\% of the real-time dataset. The output from the real-time multi-beam pipeline based on the Astro-Accelerate compatible GPU version of Pulscan will be used as the sifting metric to determine which beams and DM-trials will be recorded for subsequent processing. Pulscan will also be used to accelerate the FDAS/FDJS process by reducing the fraction of Fourier bins which need to be searched. We present the results of testing Pulscan in a synthetic environment designed to be relevant to this application in Section \ref{results_sensitivity} and Section \ref{results_execution_time}.

\section{Searching for binary pulsars}

In this section we will explore a more detailed description of the fundamental algorithms that are used in each approach.

\subsection{Matched Filtering}

\label{matched_filtering_section}

A matched filter is a template of the exact signal of interest. In the context of searching for binary pulsars, searching a dataset with matched filtering involves convolving the complex frequency spectrum of a dedispersed time-domain signal collected on a radio telescope with a range of matched filters. Each filter corresponds to a pulsar signature that has spread across a certain number of frequency bins, also referred to as $r$-bins. When there is a match between the frequency spectrum and the matched filter, there will be a spike in the response at the spin frequency of the pulsar. Essentially, the filter consolidates the power spread across multiple Fourier bins back into a single output bin. The number of bins over which the signal has drifted during the course of the observation is referred to as $z$. The parameters ($r$ and $z$) of the template which leads to a match can be used to infer details of the parameters of the orbital system that led to the creation of the signal in the frequency spectrum. Typically, there will also be a detectable response at integer multiples of $r$ and $z$ , which correspond to the higher harmonics of the signal, particularly in the case of low duty-cycle pulsars. The higher harmonics are a byproduct of the frequency decomposition of low duty-cycle periodic signals.

In a standard FDAS, the matched filters correspond to the frequency smearing pattern of a linear chirp with the form:

\begin{equation}
    s(t) = A\cos(\phi_0 + \pi(2 f_0t + ct^2))
\end{equation}

where $\phi_0$ is the initial phase, $f_0$ is the starting frequency and $c$ is the chirp rate. This creates the implicit assumption that FDAS will only lead to strong responses in cases where the pulsar signature exhibits a linear change in frequency during the observation, and closely resembles the matched filter. This linear approximation is only valid over short sections of a sinusoid, this is formalised in \cite{Ransom2001} and the constraint:
\begin{equation}
    T_{\text{obs}} \lesssim \frac{P_{\text{orb}}}{10},
\end{equation}
where $T_{\text{obs}}$ is the total integration time of the observation and $P_{\text{orb}}$ is the period of the orbit of the binary system. The parameter that varies between the templates is referred to as $z$, which is a positive or negative number that corresponds to the direction and number of frequency bins that the signal drifted between during the observation. The astronomer can use the measured value of $z$ to calculate the time derivative of frequency $\dot{f} = \frac{z}{T^2}$, which in turn leads to a physical measurement for the acceleration of the pulsar in $\mathrm{m\,s^{-2}}$ via:

\begin{equation}
    \text{Acceleration} = \frac{\dot{f}c}{f},
\end{equation}

Where $c$ is the speed of light, $f$ is the fundamental spin frequency of the pulsar and $T$ is the integration time of the observation.

A standard FDAS can be extended into a Fourier Domain Jerk Search (FDJS, \cite{Andersen2018:JERK}), where for each $z$ value, extra templates are tested that include a linear change in acceleration. The goal of an FDJS is to increase the upper bound on the $\frac{T_{obs}}{P_{orb}}$ range of sensitivity from 0.10 to 0.15, so systems with a shorter $P_{orb}$ become detectable with a fixed $T_{obs}$.

This leads to an extra dimension in which the templates can vary, referred to as $w = \dot{z}$, which is the number of $z$ bins drifted during the course of the observation. Then $w$ can be used to calculate the rate of change of $\dot{f}$, which can then be used to calculate a physical ``jerk" value for the pulsar in units of $\mathrm{m\,s^{-3}}$ using:

\begin{equation}
    \text{Jerk} = \frac{wc}{fT^3}.
\end{equation}

While an FFT-based search for approximately periodic pulsars is a 1D search across $r$ frequency bins, using FDAS extends this to a 2D search across $r$ frequency bins and $2 \times z_{max} - 1$ (to include positive and negative values of $z$) $z$-bins. A Jerk search extends this further to a 3D search across an $r, z, w$ volume. 

Since the jerk search can be thought of as consisting of multiple FDAS searches at varying values of $w$, if the astronomer wants to search a wide range of $z$ and $w$ values, or wants to perform an FDAS on a high resolution time series (large $r$) the computational cost can become impractical to perform in real time.

        \subsubsection{Matched Filtering with PRESTO}

The de-facto standard implementation of matched filtering for binary pulsar searching on CPUs is implemented in PRESTO \cite{Ransom_2003}.

PRESTO is a toolkit for processing time domain radio astronomy data, but in this work we are particularly concerned with the \code{accelsearch} program. This program implements both FDAS and FDJS, which the user can select by passing non-zero arguments to the \code{-zmax} and \code{-wmax} flags respectively.

In this work we use PRESTO as a gold standard for quantifying the sensitivity and execution time of a CPU based, FDAS/FDJS matched filtering approach.

        \subsubsection{Significance Calculation in FDAS/FDJS}

For this analysis we will consider zero-mean, unit-variance Additive White Gaussian Noise (AWGN). The complex FFT spectrum, denoted as \( \mathbf{X} \), can be formalised as follows, where \( r \) represents the number of frequency bins in the FFT spectrum:

\begin{equation}
    \begin{aligned}
        \mathbf{X} &= [x_0,\,x_1,\,\ldots,\,x_{r-1}], \\
        \text{where:} \quad x_i &= a_i + b_i j, \\
        j^2 &= -1, \\
        a &\sim \mathcal{N}(0,\,1), \\
        b &\sim \mathcal{N}(0,\,1).
    \end{aligned}
\end{equation}

Let the individual complex matched filters, denoted as \( \mathbf{M} \), where \( z \) indicates the number of Fourier bins drifted by the signal, be defined by:

\begin{equation}
    \begin{aligned}
        \mathbf{M} &= [m_0,\,m_1,\,\ldots,\,m_{z-1}], \\
        \text{where:} \quad m_i &= c_i + d_i j. 
    \end{aligned}
\end{equation}

For each combination of \( r_i \) and \( z \), a matched filter corresponding to the drift \( z \) is applied at the frequency bin \( r_i \). The result is a linear combination of samples from a Gaussian distribution:

\begin{equation}
    \begin{aligned}
        \mathbf{Y}_{i, z} &= [x_i\cdot m_0 + x_{i+1} \cdot m_1 + \ldots + x_{i+z_{max}-1} \cdot m_{z_{max}-1}].
    \end{aligned}
\end{equation}

Consequently, \( \mathbf{Y} \) remains a complex array, with both its real and imaginary components being Gaussian random variables.

When the output of FDAS/FDJS is searched for candidates, for each value \( y_{i,z} \) in the complex \( \mathbf{Y} \) array, the sum of squares of the real and imaginary components is calculated as \( |y_{i,z}|^2 \). Since the sum of squares of \( k \) independent standard normal random variables follows the chi-square distribution \( \chi^2(2) \) (with 2 degrees of freedom), one can calculate a p-value for each point using the survival function of the \( \chi^2(2) \) distribution. This p-value must be adjusted using the Bonferroni correction for the number of independent trials, as detailed in Equation (6) in \cite{Andersen2018:JERK}. The p-value can then be converted into a Gaussian equivalent sigma using the survival function for the normal distribution.

\subsection{Boxcar Filtering}

Instead of using a matched filter, Pulscan implements a unit-amplitude boxcar filter. This describes the profile of a group of frequencies appearing together with the same magnitude. 

By applying this processing to the real-valued magnitude squared of the FFT spectrum, information about phase has been discarded, so it is impossible to tell from this approach the exact nature of the frequencies that are grouped together. 

In practice, this means the astronomer cannot tell whether the pulsar is accelerating towards or away from the observer without further processing the result with a phase-sensitive technique, an example of which would be FDAS/FDJS.

\subsubsection{Magnitude v.s. Phase}

\begin{figure*}[htb]
  \makebox[\textwidth][c]{\includegraphics[width=\textwidth]{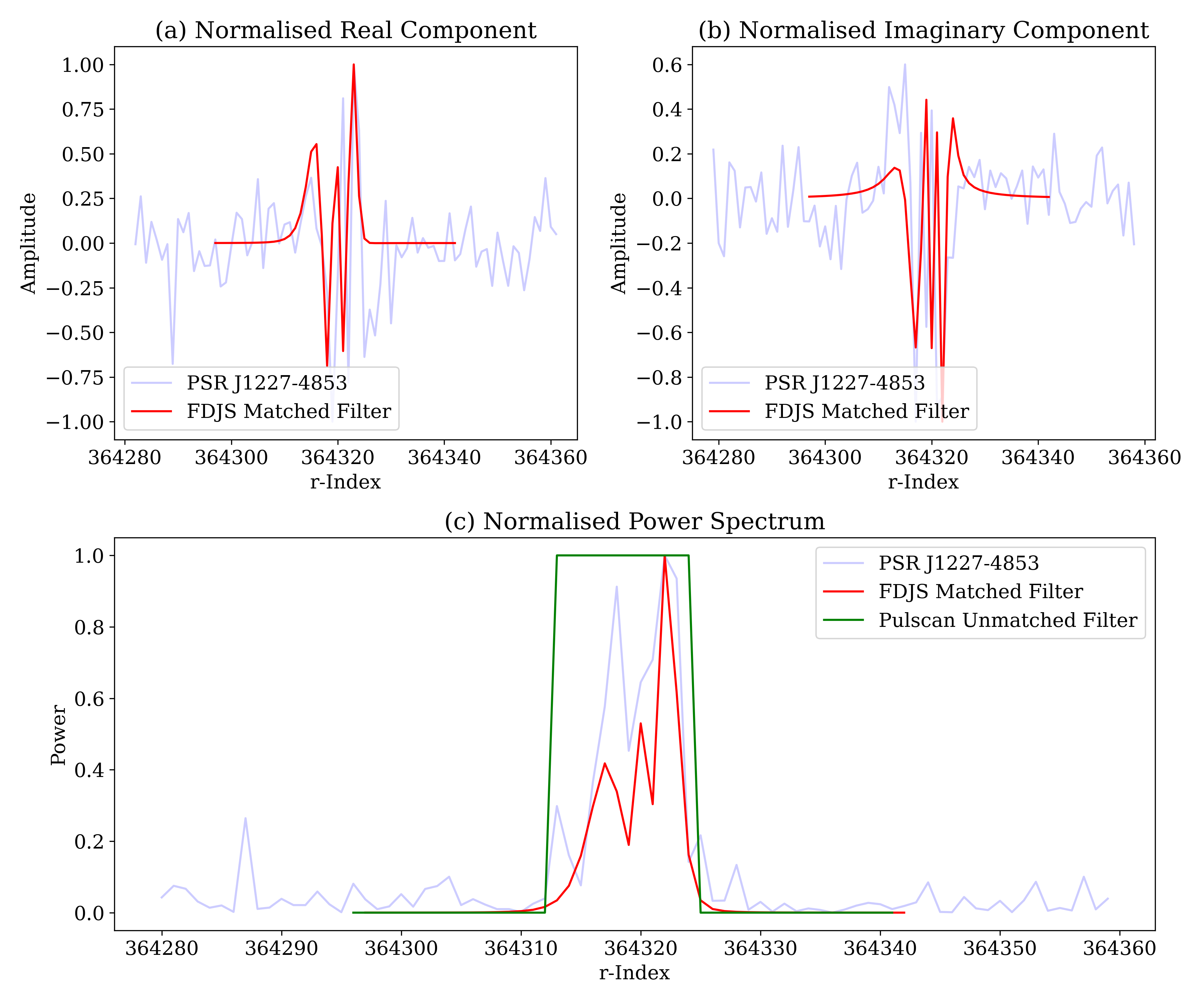}}
  \caption{Fundamental peak section of the FFT spectrum of a dedispersed observation of PSR J1227-4853, collected with the GMRT. Overlaid are the matched and unmatched filters that led to the highest significance candidate from FDJS and Pulscan respectively. }
  \label{fft_spectra}
\end{figure*}

Figure \ref{fft_spectra} shows the fundamental peak of a 1.69 ms pulsar J1227-4853 discovered at the GMRT (Roy et al. 2015). Each plot is drawn from the complex FFT spectrum of a dedispersed timeseries collected at the GMRT, the details of which are provided in Section \ref{real_data_section}. Figures \ref{fft_spectra}(a) and \ref{fft_spectra}(b) show the real and imaginary components of the complex FFT spectrum surrounding and including the fundamental peak. As described in Section \ref{matched_filtering_section}, this is the data that a complex matched filter will operate on. It can be seen that the overlaid FDJS matched filter is a good approximation to the data in both the real and imaginary components of the complex data. When the data from Figures \ref{fft_spectra}(a) and \ref{fft_spectra}(b) are combined to produce the power spectrum in Figure \ref{fft_spectra}(c), it is possible to compare the unit amplitude boxcar filter used by Pulscan with the matched filter used by FDJS. The FDJS filter is a significantly better approximation to the profile of the fundamental peak, and this is reflected in the higher significance of the detection that was made by FDJS, shown in Table \ref{tab:real_data}. However, whilst we acknowledge the sensitivity drop caused the application of boxcar filters to this detection problem, the motivation behind Pulscan is to maximize the overall number of detections for a constrained compute budget. Specifically, the goal of Pulscan is to maximize the amount of data that can be processed in real time. The benefit of the unmatched filter in Pulscan is the symmetric, uniform structure of the filter, which we have exploited using the methods described in Section \ref{execution_time_benefits} to maximize the throughput of a Pulscan-based data reduction step.

\subsubsection{Boxcar Filtering in Pulscan}

Boxcar filtering has been implemented in many software packages for many purposes, including PRESTO and AstroAccelerate where it is used to search dedispersed timeseries for single pulses, such as those caused by Fast Radio Bursts (FRBs).

Pulscan recursively computes a complete set of boxcar filtered FFT magnitude squared spectra, using the same input data as PRESTO's \code{accelsearch}, which are .fft files.

\subsubsection{Significance Calculation in Pulscan}

The input data to Pulscan is a complex FFT spectrum. Each component, real and imaginary, of the spectrum is normalised independently to ensure that they each follow a zero-mean unit-variance Gaussian distribution.

To achieve this normalisation, the input data is divided into blocks. For each block, the median serves as a robust estimator of the central location, and the median absolute deviation (MAD) is used as a robust scale estimator. However, since the MAD alone is not a consistent estimator for the standard deviation of a Gaussian distribution, it is scaled by the factor \(\frac{1}{\phi^{-1}(\frac{3}{4})}\), where \(\phi^{-1}\) is the quantile function of the standard Gaussian distribution. This adjustment ensures the transformed data has unit variance under the assumption of Gaussianity.

The formula for the normalised datapoint, after considering the aforementioned adjustments, is:

\begin{equation}
    \text{Normalised Data} = \frac{\text{Data} - \text{Median}}{\text{MAD} \times \frac{1}{\phi^{-1}(\frac{3}{4})}}.
\end{equation}

In contrast to FDAS and FDJS, the next step after normalisation in Pulscan involves calculating the square of the magnitude of the FFT spectrum. In the case of AWGN, this reduces the input data to a real-valued chi-squared distribution with \( k = 2 \) degrees of freedom for each frequency bin \( r_i \).

Following this step, \( z_{max}\) unit amplitude boxcar filters of incremental widths in the range [\(2, 3, ..., z_{max}+1 \)] are applied to the FFT magnitude spectrum. The user can select the step size to determine which boxcar filtered spectra are searched for candidates. However, all must be calculated due to the recursive implementation of the algorithm in Pulscan, detailed in Section \ref{recursive_filters}. As each boxcar-filtered output is essentially a sum of \( z_i + 1\) independent chi-squared random variables (each individually having \( k = 2 \) degrees of freedom), the output follows a chi-squared distribution with \( 2 \times (z_i+1) \) degrees of freedom:

\begin{equation}
    \begin{aligned}
        \text{Filtered Output}_{r_i, z_i} &\sim \chi^2(2 \times (z_i + 1))
    \end{aligned}
\end{equation}

To evaluate the significance of any detected peaks, one can compute a p-value for each data point using the survival function of the corresponding \( \chi^2 \) distribution. These p-values must then be adjusted using the Bonferroni correction to account for the multiple comparisons made across different boxcar widths and frequency bins. Finally, the adjusted p-values can be converted into a Gaussian-equivalent sigma, providing a normalised measure of significance for each candidate event.

\subsubsection{Execution Time Efficiency of Boxcar Filtering}
\label{execution_time_benefits}

Boxcar filtering offers computational benefits that when combined, potentially make it more time-efficient than FDAS. These advantages are derived from four primary factors:

\begin{itemize}
    \item \textbf{Computational Economy through Recursion}: The algorithmic design of boxcar filters can maximise computational efficiency by employing recursion to calculate successive filters. This optimisation minimises the total number of calculations required.

    \item \textbf{Data Volume Reduction}: Unlike FDAS, which operates on the complex Fast Fourier Transform (FFT) spectrum, the boxcar filters in Pulscan operate on the real-valued squared magnitude of the same spectrum. This effectively reduces the input data volume to the search by half.

    \item \textbf{Reduction in required number of filters} FDAS has separate (although related) filters for positive and negative accelerations, when the pulsar is accelerating towards or away from the observer. A boxcar filter based approach uses a single filter to gather the frequencies spread out by both positive and negative accelerations, reducing the number of template trials by a factor of 2.
    
    \item \textbf{Parallelisation Suitability}: Both boxcar filtering and FDAS/FDJS are well-suited for parallel implementation, particularly on Graphics Processing Units (GPUs). Prior research \citep{Dimoudi_2018} has demonstrated significant speed improvements in GPU-based implementations of FDAS, and subsequently FDJS in \citet{adamek2019jerk}. Other work designing single pulse detection codes based on boxcar filtering for GPUs \citep{adamek2020single} has also resulted in implementations that run many times faster than real time.

\end{itemize}

\subsubsection{Recursively Calculating Boxcar Filters in Pulscan}
\label{recursive_filters}

As mentioned, one of the advantages of employing boxcar filtering as opposed to matched filtering lies in the computational efficiency gained through recursive calculations. Specifically, an array filtered with a boxcar of width \( N \) can be incrementally computed by filtering with boxcar widths ranging from 2 to \( N-1 \), followed by the final boxcar filter of width \(N\).

Let \( \mathbf{X} = [x_1, x_2, \ldots, x_N] \) be the input array. Then, the unit amplitude boxcar-filtered array \( \mathbf{Y}_{k} \) for boxcar width \( k \) can be formulated as follows:

\begin{itemize}
    \item For \( k = 1 \):
    \[
    \mathbf{Y}_{1} = \mathbf{X}
    \]
    
    \item For \( k = 2 \) (Output length \( N-1 \)):
    \[
    \mathbf{Y}_{2} = [x_1 + x_2, x_2 + x_3, \ldots, x_{N-1} + x_N]
    \]
    
    \item For \( k = 3 \) (Output length \( N-2 \)):
    \[
    \mathbf{Y}_{3} = [x_1 + x_2 + x_3, x_2 + x_3 + x_4, \ldots, x_{N-2} + x_{N-1} + x_N]
    \]
\end{itemize}

Noticeably, the array \( \mathbf{Y}_{3} \) can be efficiently calculated using the array \( \mathbf{Y}_{2} \) as follows:
\[
\mathbf{Y}_{3} = \mathbf{Y}_{2}[:\text{end}-1] + \mathbf{X}[3:\text{end}]
\]

In general, the array \( \mathbf{Y}_{k} \) for boxcar width \( k \) can be recursively computed using the previous boxcar-filtered array \( \mathbf{Y}_{k-1} \) as:
\[
\mathbf{Y}_{k} = \mathbf{Y}_{k-1}[:\text{end}-1] + \mathbf{X}[k:\text{end}]
\]

This recursive relation allows for substantial computational savings, particularly beneficial for large input data and extensive searches.

This is a computational advantage over FDAS/FDJS because it is not possible with the matched filtering approach to reuse the result for a particular $r$, $z$, $w$ to calculate any other, they must all be calculated from scratch.

\subsection{Harmonic Sum}
\label{harm_sum}

To increase the detection significance of low duty-cycle pulsars where the power of the signal is spread across multiple integer harmonics of the fundamental frequency, one can use a harmonic sum on the \(r\)-\(z\) plane after performing FDAS. Successive harmonics will be detected with increasing integer multiples of \(r\) and \(z\), and can be summed accordingly. The goal is to produce a single peak with a higher significance than the individual constituent peaks corresponding to each harmonic. Since the distribution of power between the harmonics is not known a-priori, sums over different numbers of harmonics can be performed to increase the likelihood of detection.

The approach to harmonic summing implemented in Pulscan involves a decimation and summing step on the input frequency magnitude spectrum.

Formally, assuming Gaussian input, the original frequency magnitude spectrum is represented as:
\[
X_1(r_1) \sim \chi^2(2), \quad \text{for } r_1 \in \left[0, \frac{N_{\text{samp}}}{2}\right].
\]

The decimation employed in Pulscan involves summing adjacent non-overlapping pairs, triples, quartets, etc. of frequency magnitudes to produce a magnitude spectra decimated by factors of 2, 3, 4, etc. respectively.

The decimated magnitude spectrum required to align the \(H\)th peak in the harmonic spectrum with the fundamental peak would be:
\[
X_H(r_H) \sim \chi^2\left(\sum_{n=1}^{H} 2n\right), \quad \text{for } r_H \in \left[0, \frac{N_{\text{samp}}}{2H}\right],
\]
where \(\sum_{n=1}^{H} 2n = H(H+1)\) degrees of freedom.

For example, to align the second harmonic with the fundamental, we take our input frequency magnitude spectrum which is \(\chi^2\) distributed with \(k = 2\), and decimate it by a factor of 2.

We then sum the overlapping sections of each decimated spectra (aligned so the 0th bin of each spectrum is summed) and discard the non-overlapping parts. Then we proceed to boxcar filter this input as previously described, resulting in an output spectrum which is distributed following a \(\chi^2\) distribution with \(k = H(H+1)(z_i+1)\) degrees of freedom. The significance of local maxima (candidates) on these output spectra can be calculated according to this distribution.

\subsection{Approximate $r$-bin location}

Pulscan is intended to be used in conjunction with proven, highly sensitive techniques such as FDAS and FDJS. In this scenario, the goal is to use Pulscan to filter through data, ranking candidates by their significance and location on the $r$-$z$ plane. Under these circumstances it is possible to save only an approximate version of the exact frequency ($r$) location that Pulscan has identified for each candidate. The frequency spectrum is processed in blocks in all versions of Pulscan to enhance cache-compatibility by ensuring all the data that is being operated on at any given time can be stored in the smallest and fastest cache, such as the L1 cache in a CPU. By simply saving the block index (an approximate frequency location) of candidates rather than the exact r-bin they came from, it is possible to accelerate the process further. In the subsequent investigations of performance, the inclusion of this optimisation will be referred to as ``Pulscan R-Block", to reflect the approximate $r$ block index that is saved. The block width is a user defined parameter that governs the width (in $r$-bins) of the processing blocks. For the experiments in this investigation, the default value of 32768 was used for all Pulscan R-Block tests.

In a resource constrained survey, high sigma candidates from a Pulscan R-Block search could be passed to a follow up FDJS to confirm detections. The FDJS would only have to operate on a small section of the frequency spectrum equal to the block width, which would reduce the computational burden compared to a FDJS on the entire FFT spectrum.

\subsection{Median and MAD normalisation on GPUs}
\label{gpu_median_mad}

In both the CPU and hybrid CPU/GPU implementations of Pulscan, normalisation is performed on parallel chunks of the input spectrum on the CPU using the Quickselect algorithm to determine the median and MAD of each chunk. This process is trivially data-parallel by distributing chunks across CPU cores, and therefore achieves meaningful speedup when run with more OpenMP threads.

For the GPU native implementation of Pulscan, we preferred a vector parallel approach to maximise throughput and avoid running conditional, branching code on the GPU. We chose to estimate the median of a chunk of the frequency spectrum using the median of medians algorithm. This can be implemented as a parallel reduction on the GPU, where threads recursively find the median of 4 non-overlapping elements.

\subsection{Significance ($\chi^2$ $\log(p)$) calculation on GPUs}
\label{gpu_significance}

As can be seen in Section \ref{harm_sum} and Section \ref{recursive_filters}, to ensure the power distributions of candidates follow $\chi^2$ statistics, the resulting powers for high-$z$ candidates will be the sum of many constituent $\chi^2$ powers. The significance calculation is critical as it allows us to discriminate between candidates that have larger powers due to the presence of underlying signal, and those candidates which have high powers solely because they are being drawn from a $\chi^2$ distribution with a high number of degrees of freedom, and contain no signal.

In the CPU and hybrid CPU/GPU implementations of Pulscan, the cumulative distribution functions required for significance calculation are processed on the CPU after the search has been completed. To achieve higher throughput in the GPU native version, we used linear regression to fit a function to approximate the output required. This resulted in the following formula: 

\begin{equation}
    \log(p) = \frac{\text{power} \cdot f(x)}{A_1}
\end{equation}

where \( f(x) \) is defined as:

\begin{equation}
    f(x) = A_2 \times x^{16} + A_3 \times x^{15} + ... + A_{16} \times x^2 + A_{17} \times x + A_{18}
\end{equation}

and $x$ is defined as:

\begin{equation}
    x = \frac{A_1 \times \text{d.o.f.}}{\text{power}}
\end{equation}

and $A_1, ... , A_{18}$ are constants derived from the linear regression process and are provided in the Appendix.

This polynomial equation can be executed in a vector-parallel approach across all candidates on the GPU instead of the highly branched cumulative distribution functions that the CPU version uses to calculate a Gaussian-equivalent sigma value. The $\log(p)$ value can be used as an input to a pre-computed threshold that the user selects for their application. 

\section{Methodology}

In this section we will describe and justify our experimental design.

\subsection{Setting the parameter space}

We wanted to understand how Pulscan would compare to the ability of FDAS to detect highly accelerated binary pulsar systems, so we chose a parameter space that would lead to the creation of a range of millisecond pulsars in close orbits around their companion.

The distribution sampling information and parameter limits are contained in Table \ref{tab:binary_fake}.

It is important to note that the parameter range intentionally goes further than the constraint under which FDAS would be expected to maintain sensitivity, $T_{\text{obs}} < \frac{P_{\text{orb}}}{10}$. We chose a lower bound on the orbital period of 1 hour, which compared with the constant 600s observation time, demonstrates that we are violating this constraint. 

We know from \cite{pan2023binary} that systems with an orbital period shorter than this are physically plausible, and this extreme edge of the parameter space is an important one to include in our dataset.

Some combinations of parameters may not be physically realisable, but it is important that we understand how the signal recovery performance degrades when searching for systems beyond the boundaries of currently known binary pulsars.

\begin{table}[htb]
\centering
\begin{tabular}{ |c|c|c|c|  }
 \hline
 Parameter& Range& Units& Distribution\\
 \hline \hline
 Spin period&[1, 10]&ms &log-uniform\\
 Pulse width&[5, 50]&\%&log-uniform\\
 Input SNR &[0.01, 0.1]&&log-uniform\\
 Dispersion measure&[100, 1000]&cm$^{-3}$pc&log-uniform\\
 Bits per sample&8&&\\
 Number of channels&1024&&\\
 Sampling time&128&$\mu$s&\\
 Observation time&600&Seconds&\\
 First channel frequency&1550&MHz&\\
 Channel offset&-\,0.292968752&MHz&\\
 Additional flags&-binary&&\\
 Orbital period&[1, 10]&Hours&log-uniform\\
 Starting orbital phase&[0, 1.0]&&uniform\\
 Pulsar mass&[0.5, 5.0]&$M_\odot$&log-uniform\\
 Companion mass&[0.5, 5.0]&$M_\odot$&log-uniform\\
 Orbital eccentricity&[0, 1.0]&&uniform\\
 \hline
\end{tabular}
\caption{\label{tab:binary_fake}Configuration arguments for SIGPROC \code{fake} command to generate a binary pulsar signal. [a, b] represents the closed interval inclusive of the bounds. The Input SNR values are per pulse, and were chosen empirically to span the range of detectable to undetectable systems. This goal was achieved, demonstrated by the fact that only a subset of the data resulted in detectable binary pulsar signatures. Further details are provided in Section \ref{detection_significance}.}
\end{table}

To achieve sufficient covering of this high dimensional space, we aimed to generate a sample size of $N_{total} = 10,000$ synthetic pulsars. In total, we simulated $10,955$ pulsars.

For some experiments, we chose to remove extremely bright pulsars that were detectable with a periodicity search, which does not account for the frequency smearing effect. The goal of removing pulsars detectable by a periodicity search is to focus our dataset on a population of pulsars which are considered to be only potentially detectable by some form of acceleration search. Further details are provided in Section \ref{detection_significance}.
    
\subsection{Generating synthetic data}

We used the \code{fake} program in SIGPROC to generate synthetic binary pulsar signatures against a background of additive white gaussian noise (AWGN). 

It is important to note that this will generate data with a fixed pulse profile (square pulses) in both time and frequency, and amplitude defined by the \code{snrpeak} parameter.

The data lacks any undesirable sources of non-Gaussian background, such as RFI. This enables us to make repeatable measurements of detection significance as our assumptions of the $\chi ^2$ background noise that our peaks are being drawn from will be valid across the entire dataset.

We dedispersed the filterbank files using the \code{prepsubband} program in PRESTO to the corresponding DM of the synthetic pulsar. We then generated the frequency spectrum using the \code{realfft} program in PRESTO.

This gave us a series of .fft files, each of which contained a pulsar signature with a known rest frame spin frequency. This information is stored in a metadata file in the same subdirectory for subsequent processing.

\subsection{Comparing results from different pipelines}

It is essential to ensure that results from separate pipelines are being compared fairly. In this section we will cover the key considerations to ensure that our experiments did not favour one technique over another.

\subsubsection{Sensitivity}

Both pipelines generate a Gaussian equivalent sigma for each output candidate. These can be directly compared as they have been treated to ensure they represent the equivalent confidence level that the particular corresponding candidate has not arisen from a random Gaussian noise process.

To ensure that we are making a fair comparison of the ability of FDAS/FDJS and boxcar filtering to extract individual binary pulsar fundamental peaks from noise, we initially disabled harmonic summing in PRESTO using the \code{-numharm 1}, \code{-noharmremove} and \code{-noharmpolish} flags. In subsequent tests evaluating the performance of relative harmonic summing approaches, we used the \code{-numharm 4} setting in both \code{accelsearch} and Pulscan.

In the sensitivity experiments, we defined detection as the ability of a particular method to produce a candidate above a certain confidence threshold within 1\% of the fundamental harmonic frequency, i.e. the spin frequency of the pulsar we knew a-priori that was present in the file being processed. The highest sigma candidate meeting this criteria was isolated from the text-formatted output candidate lists from both \code{accelsearch} and Pulscan. Section \ref{results_sensitivity} addresses the considerations necessary to use this detection criteria meaningfully.

\subsubsection{CPU Implementation Performance}

All figures quoted are averages over 16 runs, and error bars represent the measured sample standard deviation of the execution times collected.

As we are testing multiple (CPU, CPU/GPU hybrid, GPU native) implementations of a boxcar-based acceleration search, we designed experiments to profile their performance in appropriate settings.

For the CPU version of Pulscan, we use the bash command \code{time}'s \code{real} output to time the end-to-end wall clock time of executing each program, \code{pulscan} and \code{accelsearch}. Notably, in addition to searching the data, this includes the overhead of reading the input file, normalising the data and writing the output candidate list. In the experiment where we disable the harmonic sum, we are isolating the minimum viable functionality to complete a binary pulsar search, and maximising the fraction of the execution time that is spent on the filtering operations. We focus on end-to-end speed-up as this is the most relevant figure for an astronomer who wants to determine whether the trade-offs of boxcar filtering are worth the speed-up on CPUs.

We profiled the execution time of the code over a variety of input data sizes, each of which covers various combinations of observation integration time and time sampling rate.

We also profiled the execution time of the code when we varied the \code{-zmax} parameter. We adhered to the upper limit of 1200 that is present in the PRESTO implementation of FDAS/FDJS, although there is no upper limit on \code{-zmax} in our CPU implementation of Pulscan.

For the CPU/GPU hybrid version of Pulscan, we measure the total execution time of the program using the self reported timing statements. This is to ensure we do not include a measurement of the first GPU initialisation process, that would not be incurred in an always-on real time setting.

\subsubsection{GPU Implementation Performance}

Since the GPU implementation of Pulscan is intended as a data flagger to be integrated into an existing GPU-native data pipeline (AstroAccelerate), the way we have measured its performance differs.

The GPU version does not calculate an exact Gaussian-equivalent Sigma for each candidate. This is because there is a large arithmetic overhead with calculating the relevant forward and inverse cumulative distribution functions, so instead the GPU version records the $log(p)$ value that would have been used as an input to the Gaussian equivalent sigma function. Since the GPU version is intended as a binary data flagger, these $log(p)$ values can then be compared with pre-computed threshold values and checked whether the candidate is above or below the threshold that has been chosen for detection.

Additionally, as the GPU implementation of Pulscan will be integrated into a pipeline, the code will not suffer the overhead of data transfers, as it will pick up where the previous stage in the pipeline finished with pre-prepared data in pre-allocated memory locations. Therefore we define the execution time of the GPU version as solely the execution time of the GPU kernels relevant to Pulscan, ignoring the overhead of data movement to and from the GPU. When making comparisons to the CPU/GPU hybrid version, we apply the same logic and only measure the core execution time of the Pulscan operations, and not reading or writing the relevant files.

\section{Results: Sensitivity}
\label{results_sensitivity}

In this section there are two main metrics that we are interested in quantifying for each approach:

\begin{itemize}
    \item The sensitivity of the detection method, i.e. the number of pulsar fundamentals that can be detected against background noise. A true matched filter achieves optimal signal recovery (when measured as signal-to-noise ratio) against a Gaussian noise background, so for all techniques we are measuring how well the filters being used approximate the binary pulsar signals we have synthesised.
    \item The ability of the detection method to accurately measure the physical parameters of the binary system, primarily the spin frequency ($r$-bin) and acceleration (proportional to $z$-bin) of the pulsar. Accurate measurements are beneficial, as they may lead to the ability to reduce the search space for follow up confirmation and timing measurements using other techniques of the pulsar peaks initially detected by Pulscan.
\end{itemize}

We have generated 10,955 synthetic filterbank files ($T_{obs}=600s$, $T_{samp} = 128 \mu s$), and processed the dedispersed frequency spectra separately with PRESTO \code{accelsearch} using the \code{-zmax 0} option for periodicity results, up to \code{-zmax 1200} option for FDAS results and \code{-zmax 1200} and \code{-wmax 2000} for Jerk search results. For boxcar results we have used Pulscan with up to \code{-zmax 1200}. These upper bounds were chosen as they are the upper limits in PRESTO.

Our entire dataset contained 10,955 synthetic filterbanks. As previously mentioned, due to the stochastic sampling of our binary system parameters, only a subset of these led to a binary pulsar signature that was undetectable by a periodicity search.

\subsection{Detection Significance}
\label{detection_significance}

In this section we will compare the number of pulsar fundamentals that each technique was able to detect above a Gaussian-equivalent sigma of 6 with the harmonic sum disabled. After excluding those detectable by a periodicity search above a sigma of 2 (default value in \code{accelsearch}), there were 5,751 simulated pulsars remaining in the dataset . As a reference, we performed a comprehensive FDJS using \code{-zmax 1200} and \code{-wmax 2000}. This detected the fundamental of 3,127 pulsars above a Gaussian-equivalent sigma of 6.

A detection was logged as the highest sigma candidate with a reported frequency within 1\% of the true rest frame spin frequency of the pulsar that the data contained (known a-priori). The implications of this detection criteria are explored further in Section \ref{spin_frequency_results}. We chose to vary \code{-zmax} for both FDAS and Pulscan, and measure the number of total and unique detections by each approach.

\begin{figure*}[htb]
  \makebox[\textwidth][c]{\includegraphics[width=\textwidth]{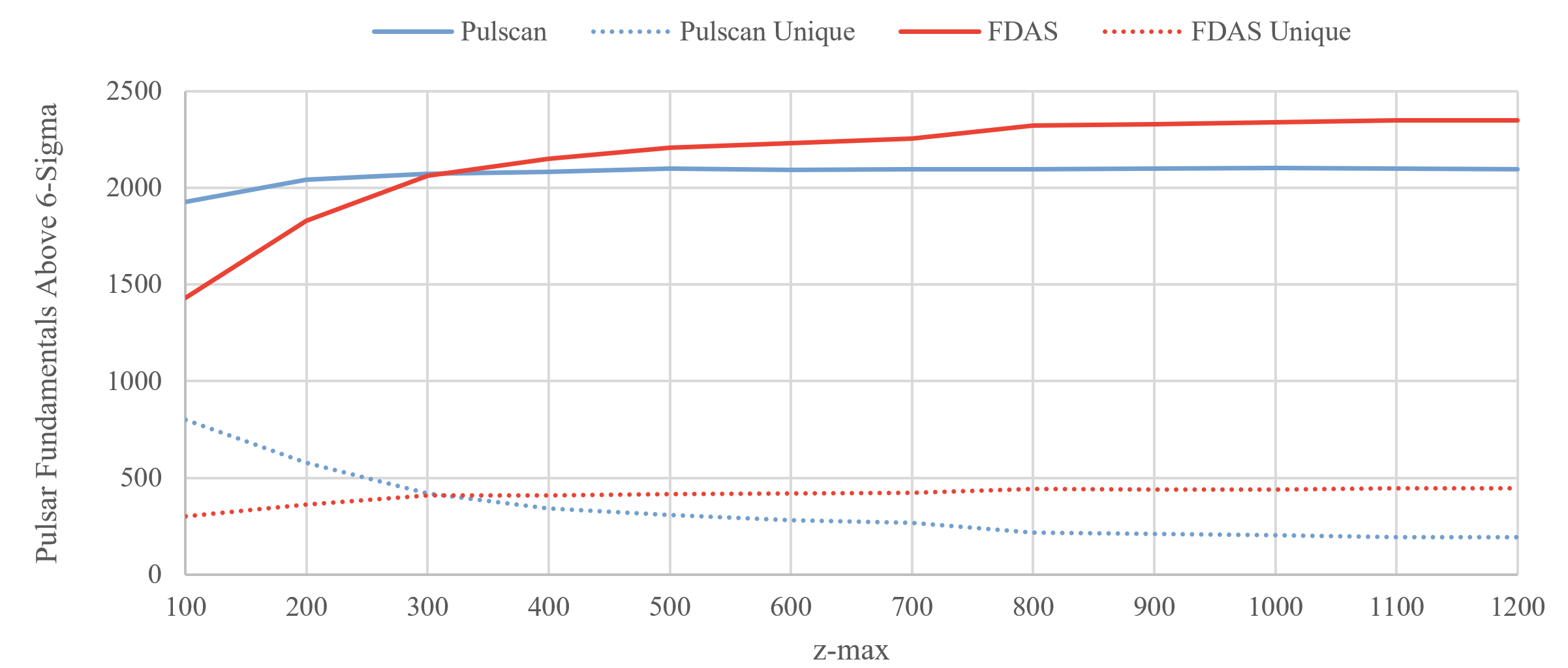}}
  \caption{Number of pulsar fundamental peaks (harmonic sum disabled) detected within 1\% of their rest frame spin frequency above the 6-Sigma threshold against varying \code{zmax} values, after removing candidates that were detectable by a periodicity (\code{zmax 0}) search using \code{accelsearch}. The ``unique" data series represent those that were not detected by the other detection method.}
  \label{z-max}
\end{figure*}

It can be seen from Figure \ref{z-max} that both FDAS and Pulscan detect fewer than 3,127 pulsar fundamentals above a sigma of 6, regardless of the \code{-zmax} setting. This demonstrates the value of the FDJS as a highly sensitive technique that will detect pulsars that are undetectable by these less computationally extensive methods. It is important to note that these results will be highly dependent on the underlying distribution of pulsar signatures in the dataset being searched. It can be seen in Figure \ref{zbin_comparison} that the comprehensive FDJS detected pulsar fundamentals spanning the full range of z-bin $0 \rightarrow 1200$, so it is expected that neither technique detected all the pulsars in the dataset when the \code{-zmax} setting was significantly under 1200. There may also be pulsars present in the dataset with a true $z$-value greater than 1200, but due to this value being the upper limit of \code{zmax} in PRESTO, we were not able to quantify the number that FDJS would detect with this experiment.

Figure \ref{z-max} shows there are two key regions in which the relative performance of each technique differs, and each region demonstrates the benefits and drawbacks of matched and unmatched filtering. 

In the \code{-zmax} $<$ 300 region, Pulscan detects more pulsar fundamentals above the sigma threshold than FDAS, and shows only a marginal improvement when the number of templates is increased by increasing the \code{-zmax} parameter from 100 to 200, and to 300. In contrast, FDAS detects fewer pulsar fundamentals, but shows a more marked increase in the number of pulsar fundamentals detected with the number of templates tested.

It is important to consider the results in the \code{-zmax} $<$ 300 region of Figure \ref{z-max} alongside the results presented in Table \ref{tab:real_data} and Figure \ref{zbin_comparison}. Table \ref{tab:real_data} demonstrates that when sufficient matched filters are used to achieve a true match between the filters being tested and the pulsar under investigation, that coherent matched filtering (either FDAS or FDJS) achieves higher signal recovery than the unmatched boxcar filters in Pulscan. The results in Figure \ref{zbin_comparison} demonstrate that FDJS detects pulsars spanning the range of $z$-magnitudes up to $z$ = 1200 in our dataset. Therefore, the results in the \code{-zmax} $<$ 300 region of Figure \ref{z-max} indicate that the only circumstance under which Pulscan detects more binary pulsars than FDAS is when FDAS is under-resourced, and unable to perform sufficient matched filter trials to match the pulsar signatures present in the data. 

We confirmed this by calculating the average magnitude of $z$ and $w$ measured in the FDJS for the 777 ``Pulscan Unique" detections at \code{-zmax 100}. The average $z$ value for the FDJS detection using \code{-zmax 1200} and \code{-wmax 2000} in cases where Pulscan made a unique detection was 396. The average $w$ value for the FDJS detection in cases where Pulscan made a unique detection was 410. This demonstrates that by limiting FDAS to a \code{-zmax 100} search with \code{-wmax 0}, we are severely under-resourcing the matched filtering approach to fully cover the underlying distribution of binary pulsars in our synthetic dataset. This is further emphasised by the rise in the number of detections corresponding to the rise in \code{-zmax} for FDAS.

However, the goal of Pulscan is to enable binary pulsar searching in resource constrained settings, and the key insight from these results is that in our experiments, unmatched boxcar filters retain acceptable sensitivity under conditions of significant undersampling of the $z$-axis to cover the pulsar distribution of our dataset. Further investigation is provided in Section \ref{log_sampling} and Section \ref{z_accuracy_experiments}.

In the \code{-zmax} $>$ 300 region, FDAS detects more pulsar fundamentals above the sigma threshold than Pulscan. This once again demonstrates the value of matched filtering in situations where there is available compute to process more filters, as they more closely match the underlying signal they are trying to detect, and lift more fundamental peaks above the 6-sigma threshold required to be classified as a detection in this experiment. Figure \ref{z-max} demonstrates that increasing \code{-zmax} in Pulscan results in diminishing returns in the high \code{-zmax} regime. 

The majority of the pulsar fundamentals detected by each technique were the same, as shown by the low overall fraction of the total detection area covered by the area under the dotted lines for the unique detections. Notably, at all \code{-zmax} values Pulscan does detect some unique pulsar fundamentals. This suggests that if the cost of running Pulscan was low enough, it could still be included as an addition to an FDAS-based pipeline, to detect the remaining pulsar fundamentals that FDAS might have missed.

\subsubsection{Unique Detections}

At the all $z$-max setting tested (up to $z$-max = 1200), it can be seen in Figure \ref{z-max} that both Pulscan and FDAS make a varying amount of unique detections. This represents the number of Pulsar fundamentals that were not detected by the other method.

\begin{figure*}[htb]
  \makebox[\textwidth][c]{\includegraphics[width=\textwidth]{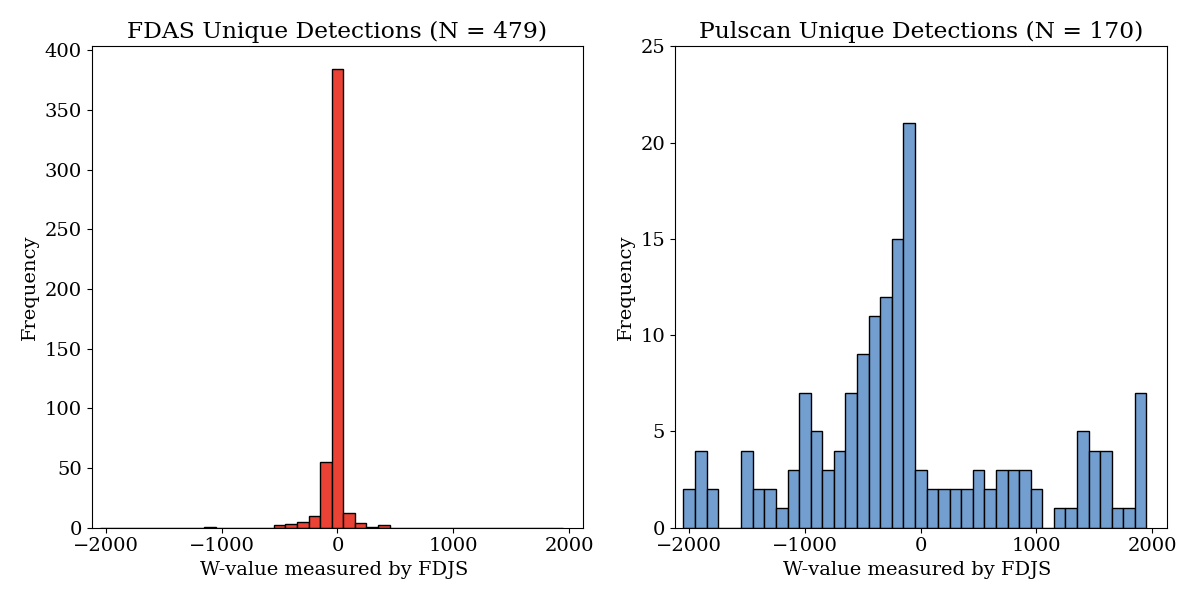}}
  \caption{Histograms of $w$-values (measured by FDJS) of the unique detections by FDAS and Pulscan when run over our synthetic dataset with a $z$-max setting of 1200. Unique detections were those that each search technique (FDAS, Pulscan) reported a candidate above 6-sigma and the other search technique did not.}
  \label{unique-detections}
\end{figure*}

To minimize the fraction of Pulsar fundamentals that were not detected due to an insufficient $z$-max setting, we focused on the unique detections at $z$-max = 1200 for both FDAS (N = 479) and Pulscan (N = 170). Using the FDJS in PRESTO, we gathered the measured $w$-value for each unique detection and plotted a histogram of the values, presented in Figure \ref{unique-detections}. 

It can be seen that the $w$-values of the unique detections of FDAS are distributed closely around $w$ = 0. This is expected and demonstrates the primary goal of performing an FDJS. In contrast, the unique detections made by Pulscan have a broader range of $w$-values, spanning the range of $-2000 < w < 2000$. This suggests that Pulscan can be used to detect jerked signals that FDAS wouldn't be able to detect, assuming the same fixed 6-sigma threshold for each approach.

When run with $z$-max = 1200 over our synthetic dataset, FDAS makes 479 unique detections in contrast to the 170 unique detections made by Pulscan.

The ability of each approach to make their own unique detections could justify running both Pulscan and FDAS in parallel in surveys where sufficient computational capacity is available.

\subsubsection{Harmonic Sum}

We re-performed this experiment by excluding any files which contained a pulsar detectable above 6 sigma by a periodicity search with harmonic sum of up to 4 harmonics enabled (\code{-zmax 0} and \code{-numharm 4}). This left 5684 pulsars.

We then ran FDAS and Pulscan over the remaining data with \code{-zmax} settings of $[50, 100, ..., 400]$ and the \code{-numharm 4} flag enabled, to finely sample the region where the two techniques were comparable in the previous experiment.

\begin{figure*}[htb]
  \makebox[\textwidth][c]{\includegraphics[width=\textwidth]{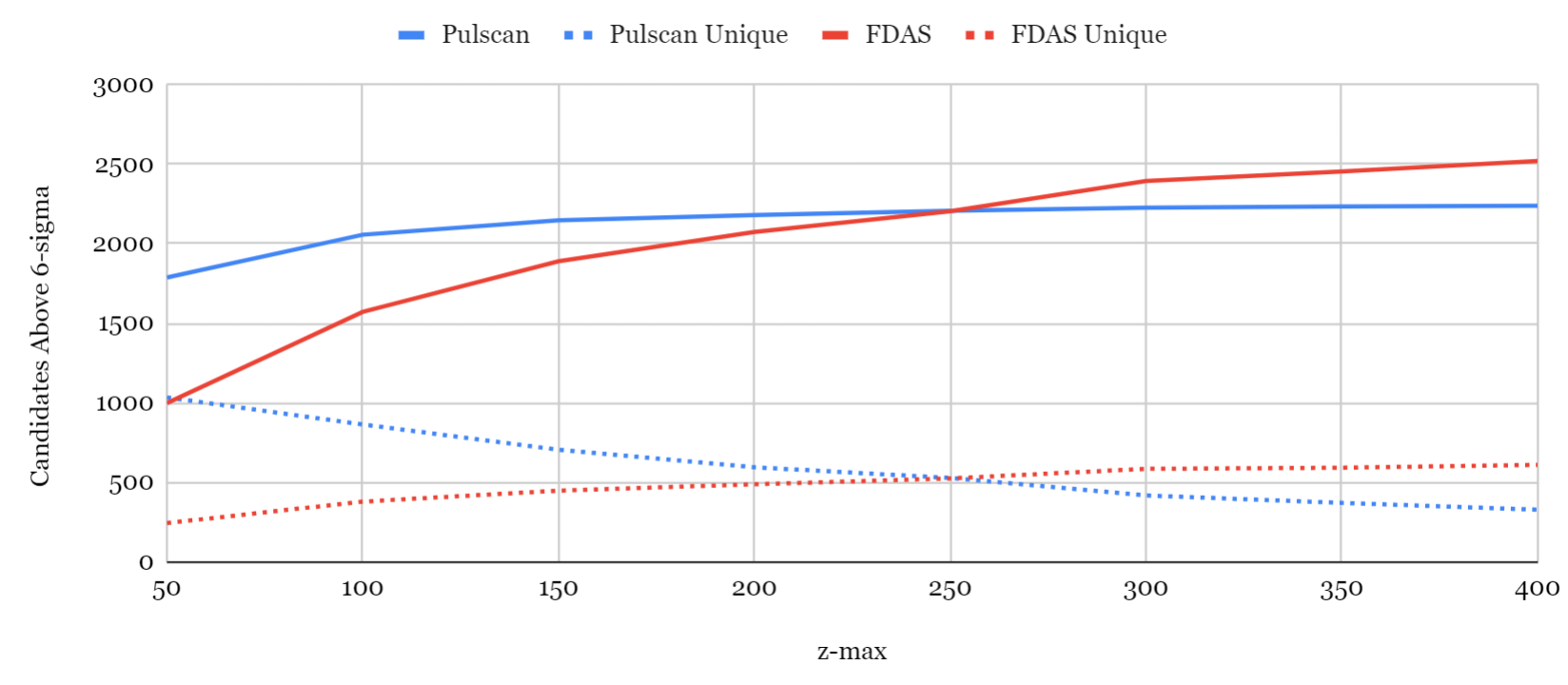}}
  \caption{Number of candidates detected (harmonic sum enabled) within 1\% of their true spin frequency above the 6-Sigma threshold against varying \code{z-max} values, with \code{-numharm 4} setting enabled in both Pulscan and FDAS, after removing any candidates detectable by a periodicity (\code{-zmax 0 -numharm 4}) search using accelsearch. The ``unique" data series represent those that were not detected by the other detection method.}
  \label{zmax_numharm_4}
\end{figure*}

Similar conclusions can be drawn from Figure \ref{zmax_numharm_4} as from Figure \ref{z-max}. At low \code{-zmax} settings, Pulscan makes more detections per filter than FDAS, however this effect does not scale with an increased number of boxcar filters. As more matched filters are tested by increasing \code{-zmax} in FDAS, the number of pulsars detected above 6-sigma continues to rise.

\subsubsection{Logarithmic Sampling of the $z$-axis}
\label{log_sampling}

Due to the recursive nature of the boxcar calculation as it is implemented in Pulscan, regardless of the $z$-max setting, it is essential for every width of boxcar up to and including $z$-max to be calculated. Notably, a significant proportion of the time is spent searching each subsequent boxcar filtered spectrum for candidates, and therefore we designed Pulscan GPU to calculate all boxcar filter widths, but only search logarithmically spaced (i.e. $z$ = 1, 2, 4, 8, ... $z$-max) boxcar filter widths for candidates. This is possible in Pulscan due to the versatility of Boxcar filters demonstrated in the previous results.

To isolate the impact on sensitivity of this change, we implemented it as an option in the CPU version of Pulscan. Then we ran Pulscan with the option both enabled and disabled over the full dataset of synthetic pulsars. 

With the logarithmic sampling option disabled, Pulscan detected 7202 candidates above a sigma threshold of 6, while detecting 7111 candidates with the option enabled. The value of this modification is a $4.1\times$ speedup (average of 8 repeats) when run with a $z$-max of 1024 and summing 4 harmonics when run on a single CPU code of an Intel Xeon Gold 6342. This speedup justified the inclusion of logarithmic z-sampling to the GPU version, as it is designed for the performance sensitive real time pipeline described in Section \ref{NSM-GMRT}.

\subsection{Pulsar Spin Frequency Measurement}
\label{spin_frequency_results}

We are  evaluating the performance of Pulscan as a data reduction technique to pass candidates to a phase-sensitive matched filtering technique such as FDJS. Firstly, it is essential to establish that there is no false positive problem and secondly that the degree of agreement between the reported r-bin location of candidates between the methods allows us to pass a reduced r-range from Pulscan to FDJS.

\subsubsection{False Positives}

The synthetic filterbanks generated for the dataset contain only Additive White Gaussian Noise (AWGN) and a single binary pulsar signature generated with square frequency and time profiles. They are not contaminated by any sources of RFI that we would experience using real data, therefore we can reasonably assume that any significant candidate produced away from the integer harmonics of the binary pulsar spin frequency is a false positive.

In the previous section, we looked for a candidate within +/- 1\% of the rest frame spin frequency of the pulsar. This is convenient, because it would correspond to a 50$\times$ reduction in the amount of data that would need to be passed to a FDJS to confirm a detection made by Pulscan.

To ensure we weren't ignoring a false positive problem, we analysed the output of performing Pulscan with \code{-zmax 1200 -numharm 1} across the entire dataset of 10,955 simulated pulsars. We blindly calculated the Mean Absolute Percentage Error (MAPE) of the frequency location of the highest sigma candidate above the 6-sigma threshold, without using our a-priori knowledge of the rest-frame frequency location of the underlying pulsar signal to select a candidate. The MAPE of the frequency location of the 3256 candidates above 6-sigma compared to the a-priori known rest frequency location was 0.936\%. There were 28 outliers with an error of between 99.5\% and 100.5\%, indicating that the first harmonic of the fundamental had been identified rather than the fundamental. The MAPE of the remaining 3228 candidates, not including the 28 outliers was 0.0771\%, and the maximum percentage error of the remaining 3228 was 1.19\%. This suggests that over our synthetic dataset, Pulscan reliably detected either the first or second harmonic as the highest sigma candidate.

\subsubsection{$r$-bin Accuracy}

After establishing that Pulscan is not generating spurious candidates at all frequencies to saturate the detection criteria with candidates, the next evaluation metric is how well the $r$-bin reported by Pulscan and FDAS for a given binary pulsar matches the $r$-bin reported by FDJS. To test this, we ran all three detection methods with harmonic summing disabled across the entire dataset of 10,955 synthetic pulsars, using \code{-zmax} = 1200 for all methods and \code{-wmax} 2000 for FDJS. We selected the 6551 pulsar signatures where all of Pulscan, FDAS and FDJS produced at least one candidate each above 6-sigma. We blindly selected the candidate with the highest significance from each of Pulscan and FDAS and compared the r-bin location with the highest significance candidate reported by FDJS. 

It is important to compare against the r-bin reported by FDJS rather than the predicted average frequency of the signature, as we are not evaluating Pulscan as a standalone detection method, and instead we want candidates from Pulscan to be a subset of those detected by a follow up method, such as FDJS.

There were 6551 cases where Pulscan, FDAS and FDJS all produced a candidate above 6 sigma.

In 6411 cases, the highest significance candidate in FDAS was within 1200 $r$-bins of the highest significance candidate in FDJS. Of these, the mean number of $r$-bins between the candidates reported by FDAS and FDJS was 7.01 and the maximum was 982. In all 140 cases where there was a discrepancy greater than 1200 $r$-bins, the discrepancy was manually verified to be a case of FDAS and FDJS detecting different harmonics of the same signal as the highest sigma candidate. In all 140 cases, the $r$-bin reported by FDAS was within 0.011\% of the frequency of a harmonic of the signal reported by FDJS.

This compares to 6373 cases where the highest significance candidate from Pulscan was within 1200 $r$-bins of the candidate produced by FDJS. The mean number of $r$-bins between the candidates reported by Pulscan and FDJS was 9.08 and the maximum was 711. In the 178 cases where there was a greater discrepancy than 1200 $r$-bins between the Pulscan and FDJS $r$-bin, the maximum relative distance of the $r$-location of a Pulscan candidate to the $r$-location of a harmonic of the FDJS candidate was 0.0482\%.

These results demonstrate that either FDAS or Pulscan could be used to reduce the size of the $r$-axis by providing candidates to be subsequently confirmed by FDJS.

\subsection{Pulsar Acceleration Measurement}
\label{z_accuracy_experiments}

\begin{figure*}[htb]
  \makebox[\textwidth][c]{\includegraphics[width=\textwidth]{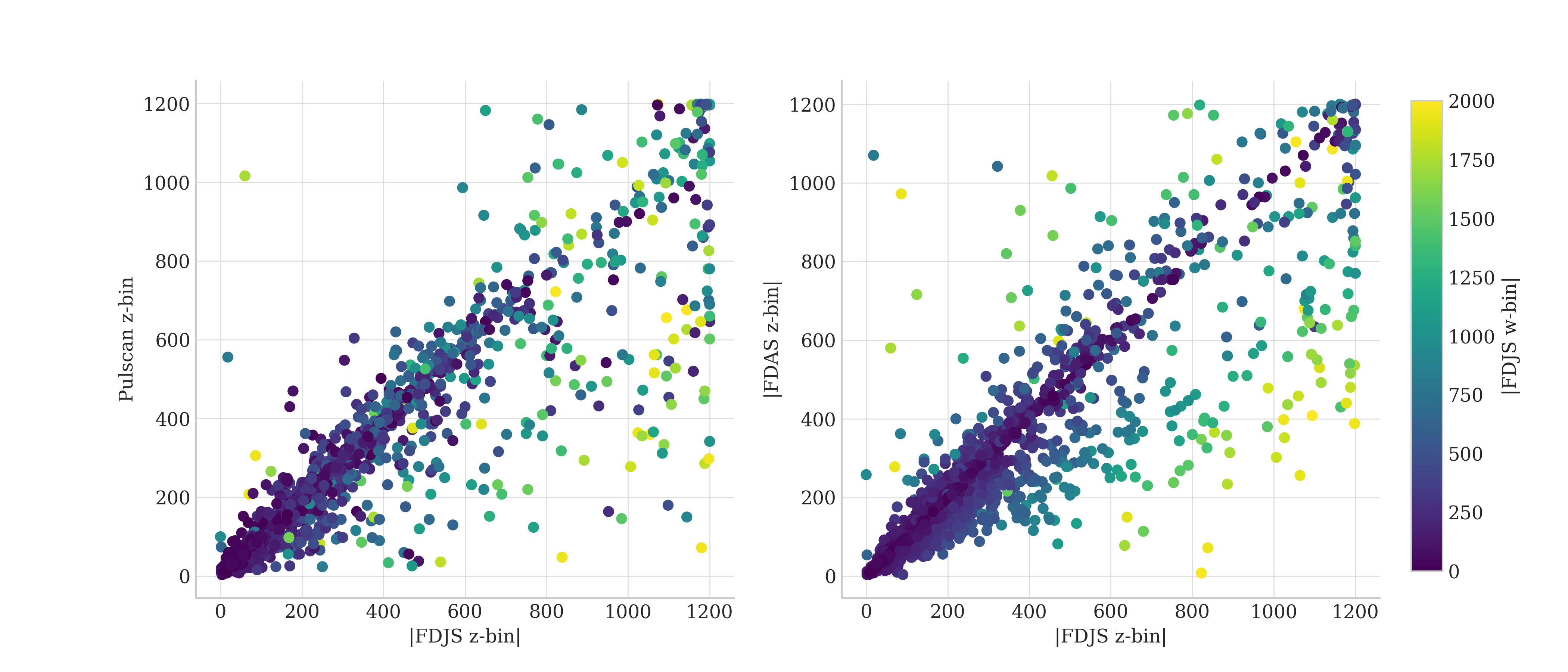}}
  \caption{Comparison of the measured \(z\)-values for Pulscan and FDAS against the reference values measured by FDJS.}
  \label{zbin_comparison}
\end{figure*}

\begin{table}[htb]
    \centering
    \caption{Comparison of Linear Regression Parameters for the \(z\)-bin measurements by Pulscan and FDAS}
    \label{tab:zbin_comparison}
    \begin{tabular}{lcc}
        \hline
        Metric & Pulscan & FDAS \\
        \hline
        Gradient & \( 0.783 \) & \( 0.852 \) \\
        Intercept & \( 23.3 \) & \( 13.9 \) \\
        \( R^2 \) Value & \( 0.827 \) & \( 0.841 \) \\
        \hline
    \end{tabular}
\end{table}

Using the same complete dataset of 10,955 datapoints processed with Pulscan, FDAS and FDJS as the previous section, Figure \ref{zbin_comparison} shows the comparison between the measured $z$-bin values from Pulscan and FDAS, compared with those measured by the FDJS.

We rely on the \(z\)-value magnitude obtained from FDJS as our reference. As before, it is important to compare the value from Pulscan to the value from FDJS as we intend to use Pulscan before using FDJS as a data reduction method.

The limitation of Pulscan to only discern magnitudes rather than signs of \(z\)-values stems from its symmetric unit-amplitude boxcar filters, which are phase-insensitive and only have width as a tunable parameter.

We assess the performance of Pulscan and FDAS based on their ability to approximate this reference value using the data in Table \ref{tab:zbin_comparison}. Neither Pulscan or FDAS perfectly recreate the results measured by FDJS. Although, FDAS demonstrates superior performance in replicating the \(z\)-values measured by FDJS, which is expected given that the set of matched filters in FDJS is a superset of those in FDAS.

In Figure \ref{zbin_comparison}, it can be seen that the degree of mismatch between the $z$-values reported by FDAS and FDJS correlates strongly with the measured value of $w$, represented by the colour scale. This shows that FDAS is more likely to overestimate or underestimate the $z$ value in the presence of highly jerked signals. In the Pulscan results, the highly jerked signals are distributed over the chart area, demonstrating that the accuracy of a Pulscan $z$-bin measurement is not strongly dependent on the jerk level of the underlying signal.

Neither Pulscan or FDAS demonstrate the ability to reduce the range of $z$ with a high degree of confidence for a subsequent FDJS around a candidate.

\subsection{Real Data}

\label{real_data_section}

To validate our findings using actual observations, we analysed data from PSR J1227-4853, a known pulsar with a spin frequency of 592.988 Hz and an orbital period of 0.2788 days. Our dataset comprises an observation of 614.4 seconds duration, equivalent to 60,000,000 samples. This data was obtained from the GMRT with a sampling time, \( T_{\text{samp}} \), of 10.24\(\mu\)s over an observing frequency span of 550$-$750 MHz.

The initial processing involved dedispersion to a Dispersion Measure (DM) of 43.4 using PRESTO's \code{prepsubband}. Subsequent steps included the removal of powerline interference with \code{zapbirds}, and a dereddening step using \code{rednoise}.

The \code{.fft} file generated was processed using FDAS in PRESTO and Pulscan.

The top candidates identified by each search method are presented in Table \ref{tab:real_data}. The parameters listed in the table align with the fundamental peak of PSR J1227-4853. The results highlight that all the employed techniques list the significant binary pulsar signal as the highest sigma candidate in the output file.

\begin{table*}[ht]
\centering
\caption{Highest significance candidate identified by each search method on real data from the GMRT. The CPU being tested was an Intel Xeon Gold 6342 and the GPU was an NVIDIA H100 PCIe.}
\label{tab:real_data}
\begin{tabular}{ccccccccc}
\hline
Technique & Hardware & \texttt{zmax} & \texttt{wmax} & Significance & $r$-bin & $z$-bin & $w$-bin & Execution Time (s)\\
\hline
Pulscan GPU&GPU&256&N/A&$log(p) = -171.341$&364288&8&N/A&0.015*\\
Pulscan Hybrid&CPU (24c) + GPU&256&N/A& $\sigma = 18.795$ & 364318 & 12 & N/A & 1.182\\
Pulscan R-Block&CPU (24c)&256&N/A& $\sigma = 18.795$ & 360448 & 12 & N/A & 2.123\\
Pulscan&CPU (24c)&256&N/A& $\sigma = 18.795$ & 364318 & 12 & N/A & 2.328 \\
Pulscan R-Block&CPU (1c)&256&N/A& $\sigma = 18.795$ & 360448 & 12 & N/A & 4.75\\
Pulscan&CPU (1c)&256&N/A& $\sigma = 18.795$ & 364318 & 12 & N/A & 14.046 \\
FDAS &CPU (1c)&256&N/A & $\sigma = 20.53$ & 364319 & -9 & N/A &28.291\\
FDJS &CPU (1c)&256&20 & $\sigma = 20.76$ & 364319 & -10 & -10 & N/A\\
FDJS &CPU (1c)&1200&2000 & N/A & 364319 & -10 & -10 & N/A\\

\hline
\multicolumn{9}{c}{*Execution time not including the time taken to transfer the data to and from the GPU.}

\end{tabular}
\end{table*}

\section{Results: Execution Time}
\label{results_execution_time}

In this section we are primarily concerned with establishing the relative execution times of each approach. We present how long it takes to perform a variety of searches on various hardware configurations. This is key when planning surveys as it will govern the boundaries of the search space that can be performed in time allocated on a shared cluster, or in real time on a dedicated cluster.

We perform separate investigations to benchmark the performance of Pulscan on CPU-based hardware and in GPU-accelerated environments.

Based on the relative balance between data production rate and available facilities to process the data, Pulscan is targeting situations where the data production rate is too high to store and subsequently do an archival search.

In these settings, there is evidently a strict requirement to process the data in less than real time, and we envisage astronomers involved in these applications will be highly motivated to reduce the execution time of their desired search to less than the observation time collected. 

There are four primary benefits to a faster search technique.

\begin{itemize}
    \item It can be run on more data (more dispersion measures, higher resolution time series) in a real time setting.
    \item A more extensive search can be performed in real time (such as higher z-max).
    \item The upfront cost of hardware required for a planned survey can be reduced.
    \item The ongoing environmental and financial cost (due to the energy usage of HPC facilities) of running a given survey can be reduced.
\end{itemize}

\subsection{FDAS vs Pulscan Execution Time}

The performance measurements for the CPU versions of Pulscan and PRESTO FDAS were conducted serially on a high-end workstation equipped with an AMD Ryzen Threadripper PRO 3995WX 64-core CPU, 512GB of RAM, and PCIe SSD storage. This enabled a comprehensive assessment of their capabilities in a resource-intensive setting.

In this section, we present a selection of experiments that profile the execution time of each approach.

\begin{figure*}[htb]
  \makebox[\textwidth][c]{\includegraphics[width=\textwidth]{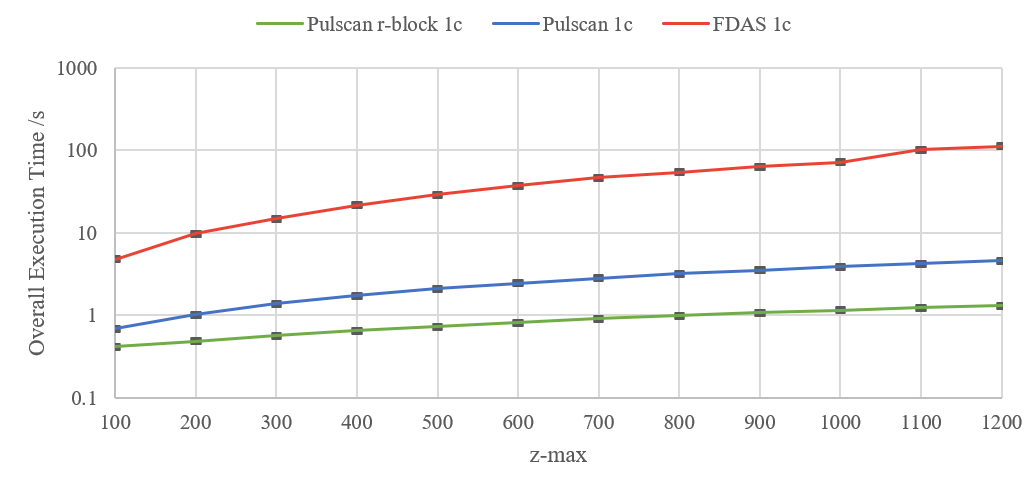}}
  \caption{Comparison of total end-to-end execution time on a single CPU core of Pulscan and PRESTO FDAS (accelsearch) against various \code{-zmax} settings. The execution time was that required to process the same input data file, a PRESTO-compatible \code{.fft} file with 8388608 ($2^{23}$) samples. Error bars represent sample standard deviation over 16 repeats.}
  \label{zmax_perf}
\end{figure*}

The results of the first experiment in Figure \ref{zmax_perf} profile the execution time of Pulscan and PRESTO's implementation of FDAS under changing the \code{-zmax} setting on a single CPU core. This parameter represents the highest level of Fourier bin drift that the search is aiming to find over the course of the observation. 

At zmax = 100, which is the least extensive search setting profiled, Pulscan using the approximate R-Block optimisation took on average 0.42 seconds to process the data, disabling the R-Block optimisation increased the execution time to 0.69 seconds. For the same \code{-zmax} setting, PRESTO's implementation of FDAS took 4.8 seconds.

At zmax = 1200, the most extensive search possible with PRESTO's implementation of FDAS,  Pulscan took on average 1.32 seconds with the R-Block optimisation enabled, and 4.61 seconds with the optimisation disabled. PRESTO's implementation of FDAS took 112.8 seconds using the same \code{zmax} setting. This demonstrates the reduced computational cost of searching extra boxcar filters by using a recursive algorithm. Figure \ref{zbin_comparison} demonstrates the value of being able to search up to very high z-max values when measuring the parameters of highly accelerated systems.

At low \code{-zmax} values, both techniques are bottlenecked by the non-filtering operations such as loading the file and normalising the data. These operations have to be done once regardless of the extent of the search and so put a minimum execution time on each approach, when the operations are run on a CPU.

\begin{figure*}[htb]
  \makebox[\textwidth][c]{\includegraphics[width=\textwidth]{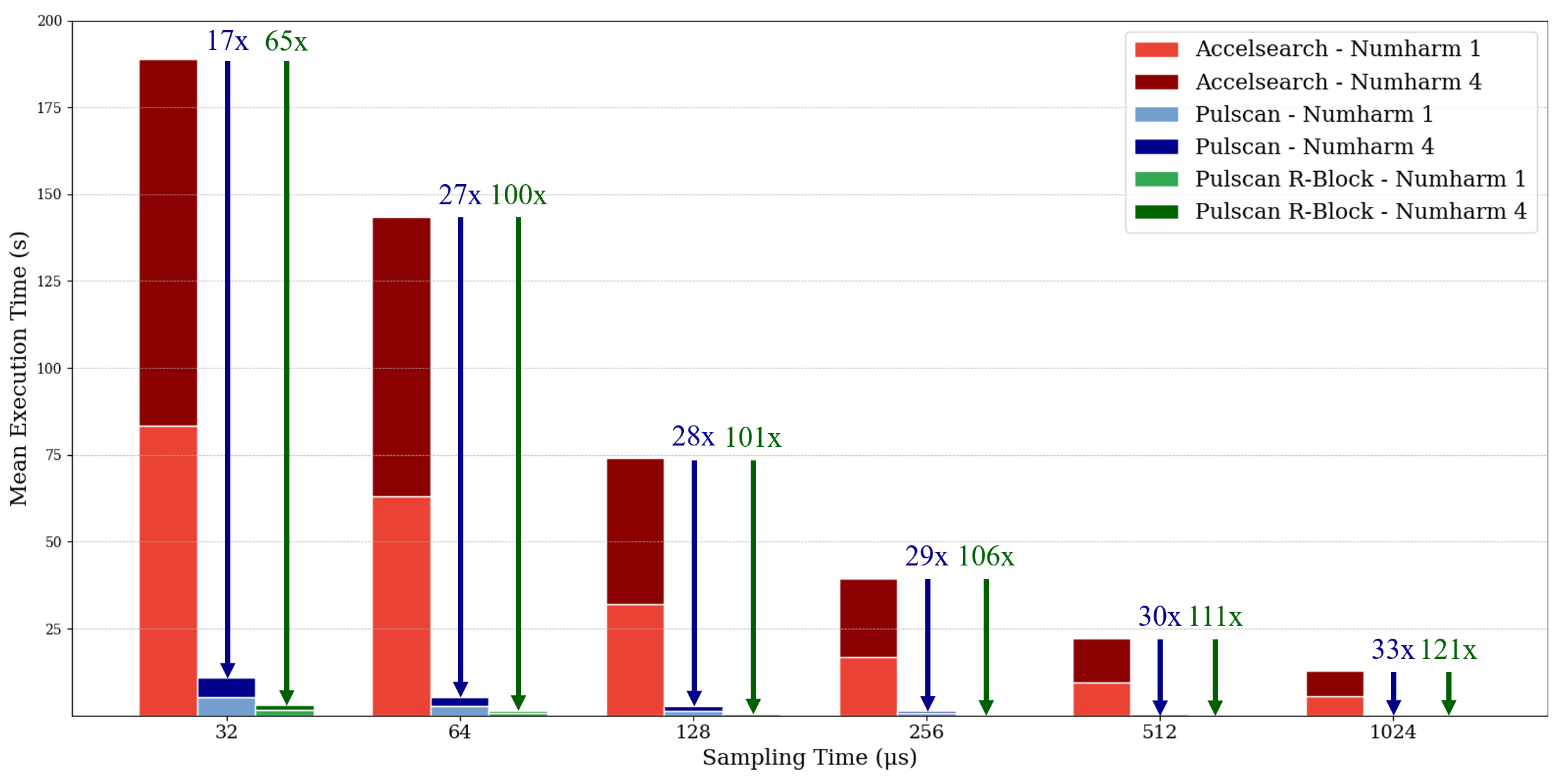}}
\caption{Comparison of end-to-end execution times of Pulscan and FDAS (accelsearch) on a single CPU core (AMD 3995WX) across different input data sizes. We generated a synthetic filterbank ($T_{obs} = 600s$) with varying $T_{samp}$ values, followed by dedispersion and Fourier transformation via PRESTO to produce \code{.fft} files. These files were subsequently analysed using both accelsearch and Pulscan, performing a search with $\code{zmax} = 1200$. The reported execution times are average values over 16 trials.}
  \label{tobs_perf}
\end{figure*}

The second experiment, presented in Figure \ref{tobs_perf}, was to vary the input data size, and measure the end to end execution time of each approach with a zmax = 1200 search on a single CPU core. Pulscan maintained a $>10\times$ speedup over PRESTO's implementation of FDAS across the entire range of input sizes tested.

\begin{figure}[htb]
    \includegraphics[width=\columnwidth]{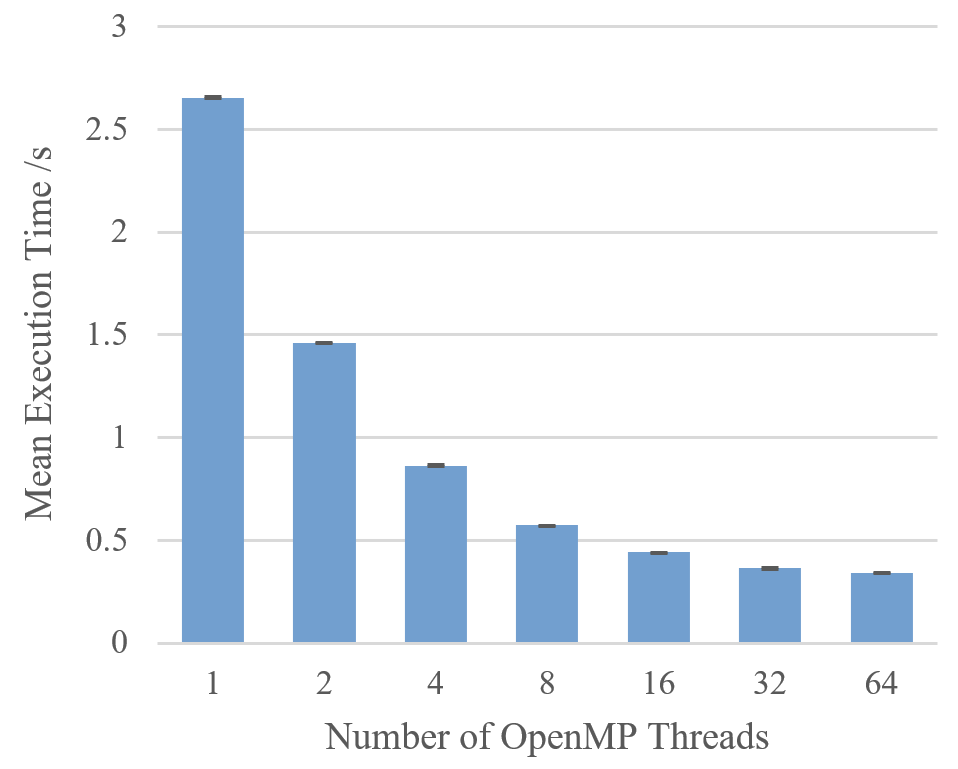}
    \caption{Comparison of end-to-end execution time of the CPU implementation of Pulscan against various \code{-ncpus} settings. The execution time was that required to perform a zmax = 1200 search with 4 harmonics summed on a PRESTO-compatible \code{.fft} file, derived from a 600s $T_{obs}$ filterbank, with $T_{samp}$ of \( 128 \mu s \). Error bars represent sample standard deviation over 16 repeats. The CPU being tested was an AMD Ryzen Threadripper Pro 3995WX.}
    \label{ncpus_perf}
\end{figure}

Both Pulscan and PRESTO's \code{accelsearch} employ OpenMP for runtime parallelisation. As demonstrated in Figure \ref{ncpus_perf}, Pulscan demonstrates notable improvement as the number of CPU threads increases. 

Due to a known bug in the code, no measurable speed-up was observed for FDAS and FDJS when using the \code{-ncpus} option. This issue isn't critical, considering that processing single \code{.fft} files in isolation is not a typical use-case for surveys, and the inherent data parallelism of processing one DM-trial per CPU core can be leveraged as a workaround.

\subsection{GPU Acceleration of Pulscan}

In this section, we will compare the performance of our implementations of Pulscan on different hardware. 

The first GPU accelerated version of Pulscan has been referred to as the Hybrid CPU/GPU version. This is because parts of the pipeline are processed on the CPU, whilst the core boxcar filtering section is performed on the GPU. The benefit of this approach is that it recreates the CPU version of Pulscan, with a true Gaussian sigma significance value calculated, and normalisation using exact median calculations. In the CPU-only version of Pulscan, the bottleneck on performance is the boxcar filtering step, and so this was our first target to offload to the GPU.

\begin{figure*}[htb]
  \makebox[\textwidth][c]{\includegraphics[width=\textwidth]{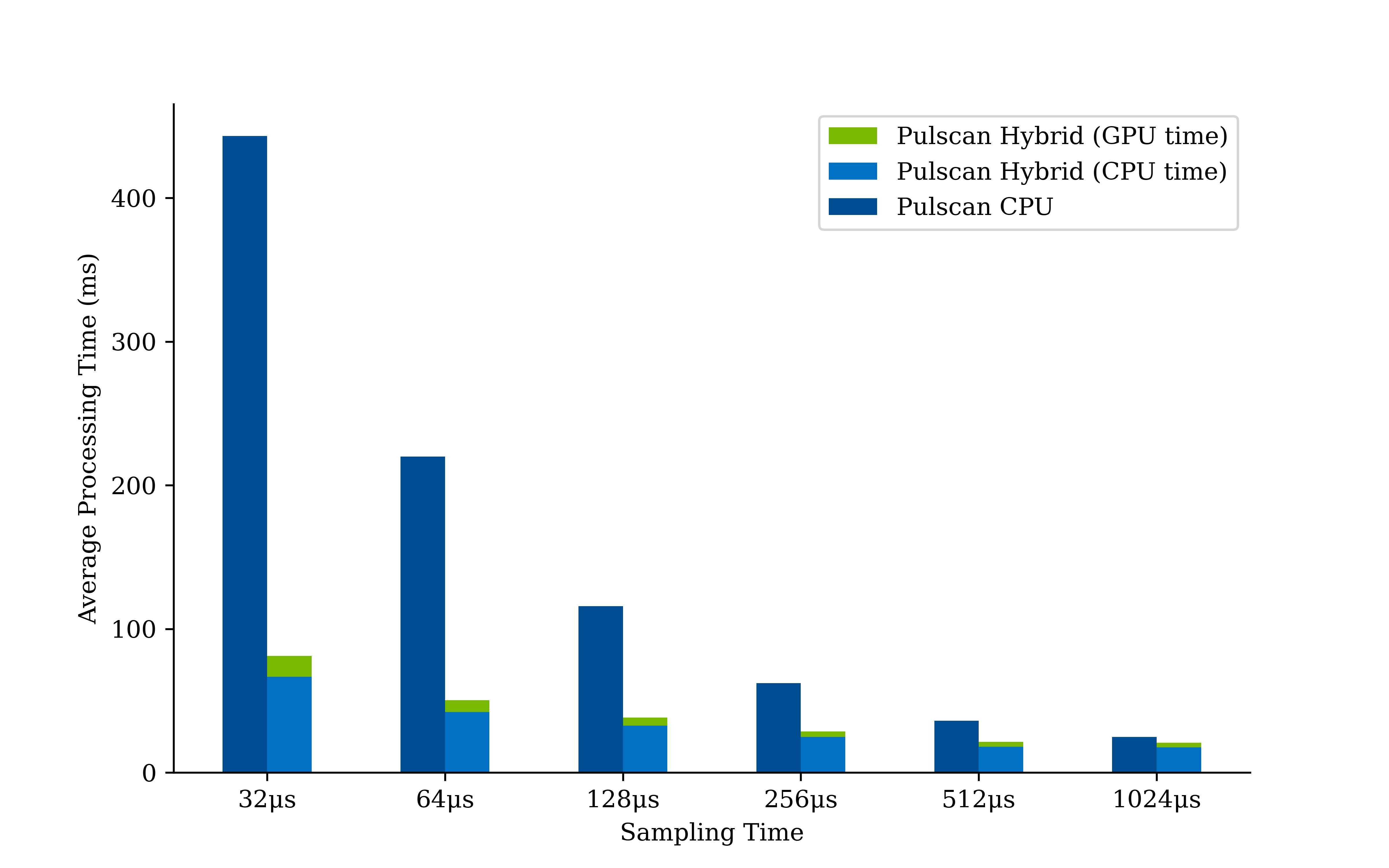}}
\caption{Comparison of processing times of Pulscan CPU vs Pulscan Hybrid CPU/GPU. The boxcar filtering was offloaded to the GPU in Pulscan Hybrid. The CPU being tested was an Intel Xeon Gold 6342 (using 24 OpenMP threads) and the GPU was an NVIDIA H100 PCIe. We generated a synthetic filterbank ($T_{obs} = 600s$) with varying $T_{samp}$ values, followed by dedispersion and Fourier transformation via PRESTO to produce \code{.fft} files. These files were subsequently analysed using Pulscan, performing a search with $\code{zmax} = 256$ and $\code{numharm} = 1$. The reported processing times are average values over 16 trials, excluding the time taken to read the data into memory and write the output to disk.}
  \label{cpu_hybrid_perf}
\end{figure*}

The results in Figure \ref{cpu_hybrid_perf} demonstrate the value of this approach on larger datasets (those with short sampling times at $T_{obs} = 600$). As can be seen, the time spent searching on the GPU version is a small fraction of the overall execution time. This leads to a $5.5\times$ overall speedup of the processing of the data at a sampling time of $32 \micro s$. The graph also demonstrates how the benefit of a hybrid CPU/GPU approach diminishes on smaller datasets. At a sampling time of $1024 \micro s$, there is only a $1.2\times$ speedup when switching from the CPU-only code to the Hybrid CPU/GPU implementation. This is an example of Amdahl's law, demonstrating that as the high latency serial CPU operations become the dominant cost of the overall execution time, the benefit of parallelising only a section of the pipeline diminishes. 

This motivates the development of a fully GPU-native pipeline, which maximises utilisation of the available GPU hardware. By using the techniques described in Section \ref{gpu_significance} and Section \ref{gpu_median_mad} (with algorithms detailed in the appendix), Pulscan GPU is able to run all operations on the GPU. This minimises the overhead of high latency serial operations which run on the CPU. Also, as Pulscan GPU is designed to be integrated into AstroAccelerate, which is a fully GPU-based pipeline, the GPU version will not have to wait for data to be transferred from the CPU memory to the GPU memory. As before, for the following benchmarks, we exclude the execution time of loading the input data from disk to memory. Further, as the output of Pulscan GPU will be passed to another section of the pipeline which also runs on the GPU, we can exclude the execution time of writing the candidates to disk.

\begin{figure*}[htb]
  \makebox[\textwidth][c]{\includegraphics[width=\textwidth]{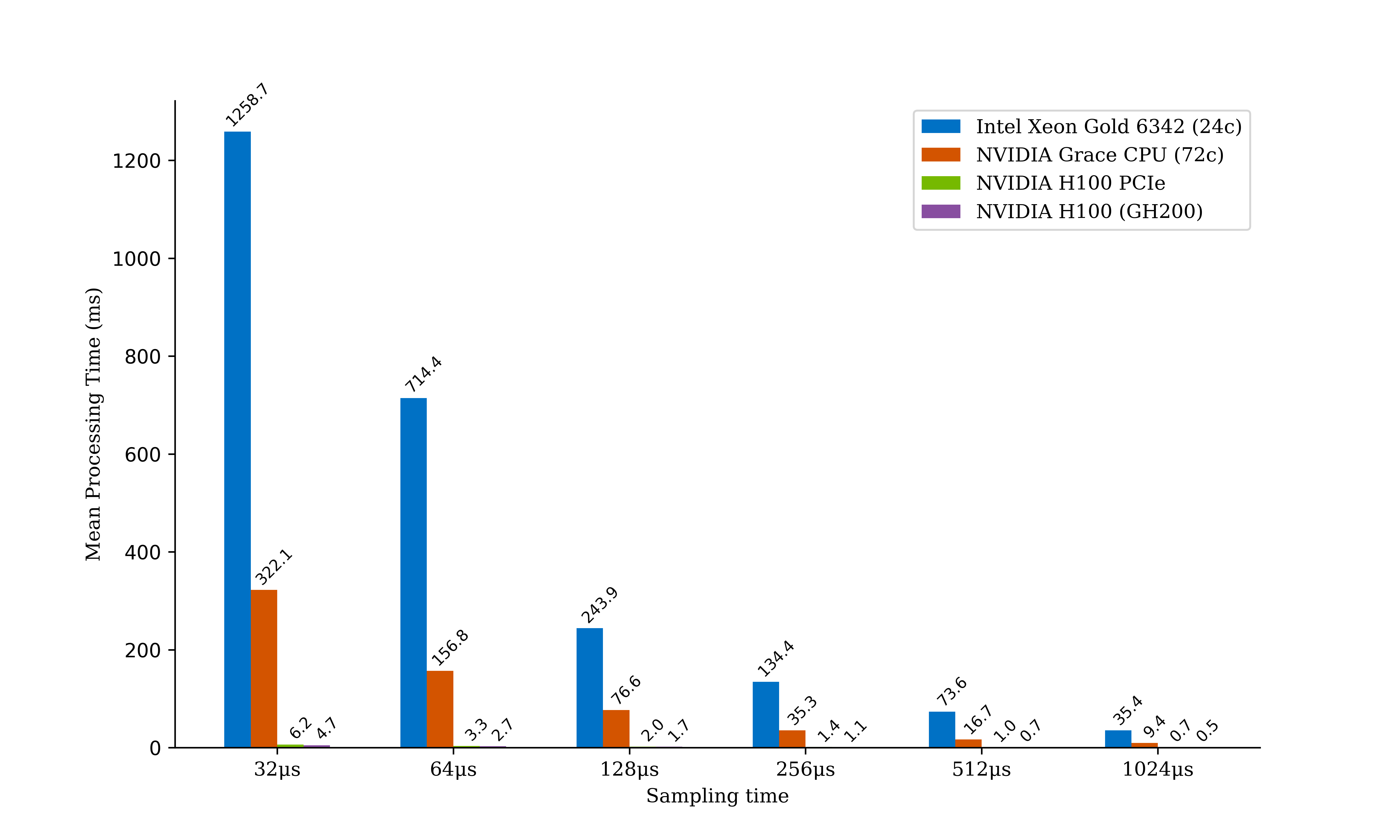}}
\caption{ Comparison of processing times of Pulscan CPU on an Intel Xeon Gold 6342 (using 24 OpenMP threads) vs Pulscan CPU on an NVIDIA Grace CPU (using 72 OpenMP threads) vs Pulscan GPU on NVIDIA H100 PCIe vs Pulscan GPU on NVIDIA H100 (GH200) across different input data sizes. We generated a synthetic filterbank ($T_{obs} = 600s$) with varying $T_{samp}$ values, followed by dedispersion and Fourier transformation via PRESTO to produce \code{.fft} files. These files were subsequently analysed using Pulscan, performing a search with $\code{zmax} = 256$, $\code{numharm} = 4$ and $\code{chunkwidth} = 256$. The $z$-axis was logarithmically sampled in both the CPU and GPU version. The reported processing times are average values over 8 trials, excluding the time taken to read the data into memory and write the output to disk.}

  \label{cpu_gpu_perf}
\end{figure*}

\begin{table*}[ht]
\centering
\caption{Performance metrics of relevant Pulscan GPU kernels reported by NVIDIA NSight Compute on an NVIDIA GH200 Grace Hopper Superchip when processing a $T_{\text{obs}} = 600\text{s}$, $T_{\text{samp}} = 128 \micro\text{s}$ observation frequency spectrum with $z$-max = 256. When maximising overall throughput on the GPU, it is important that the majority of execution time is spent on kernels with high utilisation of available compute and memory resources.}
\label{tab:kernel_performance}
\begin{tabular}{cccc}
\hline
Kernel & Duration (µs) & Compute Throughput (\%) & Memory Throughput (\%) \\
\hline
SeparateComplexComponents & 11.46 & 20.03 & 40.85 \\
MedianOfMediansNormalisation & 14.4 & 38.61 & 39.67 \\
MagnitudeSquared & 11.3 & 22.37 & 40.24 \\
DecimateHarmonics & 9.31 & 36.63 & 47.19 \\
BoxcarFilterArray & 420.26 & 79.34 & 83.35 \\
CalculateLogP & 23.26 & 51.7 & 2.91 \\
\hline
\end{tabular}
\end{table*}

Figure \ref{cpu_gpu_perf} demonstrates the increased speedup achieved by performing all operations on the GPU. Both the CPU and GPU version of Pulscan perform a boxcar acceleration search with median based normalisation, $4$ harmonics summed up to a $z$-max of $256$ and significance calculation, but the NVIDIA H100 (GH200) GPU implementation is between $70\times$ and $250\times$ faster than Pulscan CPU with 24 OpenMP threads on an Intel Xeon Gold 6342. This is a result of the combination of higher memory bandwidth being available on the GPU, higher computational throughput (FLOP/s) and GPU-optimised algorithms we designed for normalising the data, performing the boxcar search and calculating the significance of candidates. The key utilisation statistics of the GPU version are detailed in Table \ref{tab:kernel_performance}. The longest executing kernel, corresponding to the Boxcar Filtering search step achieves high (approximately 80\%) utilisation of the GPU, demonstrating how effectively the algorithm maps to the GPU hardware. The other kernels show moderate utilisation, but account for a smaller fraction of the overall pipeline execution time.

\section{Conclusions}

The analysis presented in this work helps to understand the performance and capabilities of an accelerated pulsar search based on boxcar filtering. The following points are the key summaries from each section.

\begin{enumerate}
    \item \textbf{Sensitivity and Detection:} The Fourier Domain Jerk Search (FDJS) demonstrates the value of a matched filtering approach, achieving greater sensitivity by detecting more pulsars above a given Gaussian-equivalent sigma compared to the other methods tested. However, the execution time of FDJS is significantly greater than that of FDAS or Pulscan, which motivates the development of a hybrid approach in resource constrained applications.

    \item \textbf{Pulsar Parameter Measurement:} All three techniques - Pulscan, FDAS, and FDJS - have demonstrated reliable accuracy in measuring the fundamental spin frequency of an accelerated pulsar when a detection is made. FDAS outperforms Pulscan in approximating the reference $z$-bin values obtained from FDJS.

    \item \textbf{Real Data Validation:} The real-world data from the GMRT of a compact binary pulsar J1227-4853 reaffirmed the utility of all techniques and gives an indication that our findings on synthetic data will generalise to real data. Every method detected the significant binary pulsar signal as the candidate with the highest significance.

    \item \textbf{Execution Time:} In all tests, across all configurations, Pulscan takes less time to execute than the implementation of FDAS in PRESTO when run with equivalent settings. The advantage becomes greater as a greater fraction of the overall processing time is spent on the search, rather than pre or post processing.

    \item \textbf{GPU Acceleration:} The parallel nature of boxcar filtering can be exploited by implementation on GPUs, showing further speedups.
    
\end{enumerate}

In conclusion, this study provides a comparison of FDAS and Pulscan, emphasizing the strengths and limitations of each. The findings underscore the importance of selecting the right approach based on the specific goals and constraints of a given astronomical survey.

\section{Acknowledgements}
WA and KA would like to acknowledge support received from STFC Grant (ST/W001969/1). JW would like to acknowledge support from EPSRC (2595728). The authors would like to acknowledge the use of the University of Oxford Advanced Research Computing (ARC) facility in carrying out this work \citep{richards_2015_22558}.

The National Radio Astronomy Observatory is a facility of the National Science Foundation operated under cooperative agreement by Associated Universities, Inc. SMR is a CIFAR Fellow and is supported by the NSF Physics Frontiers Center awards 1430284 and 2020265.

\bibliography{bibliography}{}
\bibliographystyle{aasjournal}

\appendix

\section*{GPU Median + MAD Normalisation}

\begin{algorithm}[H]
\DontPrintSemicolon
\SetKwFunction{FMain}{local\_median}
\SetKwProg{Fn}{Device function}{:}{}
  \Fn{\FMain{$a, b, c, d$}}{
        $min \gets \min(\min(\min(a, b), c), d)$\;
        $max \gets \max(\max(\max(a, b), c), d)$\;
        $median \gets (a + b + c + d - min - max) / 2$\;
        \KwRet $median$\;
  }

\SetKwFunction{FGlobal}{median\_of\_medians}
\SetKwProg{Fn}{Global function}{:}{\KwRet}
  \Fn{\FGlobal{$data\_array$}}{
        \textbf{shared} $sdata\_array[4 \times \text{blockDim}]$\;
        \tcp{load sdata\_array from data\_array}
        \tcp{then reduce over sdata\_array}
        \For{strides $\leftarrow$ blockDim, blockDim/4, \ldots, 1}{
            \If{threadIdx $<$ stride}{
                $a \gets sdata[\text{threadIdx}]$\;
                $b \gets sdata[\text{threadIdx} + stride]$\;
                $c \gets sdata[\text{threadIdx} + 2 \times stride]$\;
                $d \gets sdata[\text{threadIdx} + 3 \times stride]$\;
                $sdata[\text{threadIdx}] \gets \text{local\_median}(a, b, c, d)$\;
            }
        }
        \KwRet $sdata[0]$\;
  }

\SetKwProg{Pn}{Procedure}{:}{}
  \Pn{}{
    \ForEach{chunk of spectrum}{
        $median \gets \text{median\_of\_medians}(chunk)$\;
        $MAD \gets \text{median\_of\_medians}(chunk - median)$\;
        $scale\_factor \gets 1.4826$\;
        $normalised\_chunk \gets (chunk - median) / (scale\_factor \times MAD)$\;
    }
  }
\caption{Median of Medians Normalisation}
\end{algorithm}

\section*{GPU Harmonic Sum Decimation}

\begin{algorithm}[H]
\DontPrintSemicolon
\SetKwProg{Fn}{Global function}{:}{\KwRet}
  \Fn{decimation(data\_len, in\_data\_array, dec\_by\_2\_array, dec\_by\_3\_array, dec\_by\_4\_array)}{
        float $1a, 2a, 2b, 3a, 3b, 3c, 4a, 4b, 4c, 4d$\;
        $1a \gets in\_data\_array[\text{threadIdx}]$\;
        $2a \gets in\_data\_array[2 \times \text{threadIdx}]$\;
        $2b \gets in\_data\_array[2 \times \text{threadIdx} + 1]$\;
        \tcp{Assignments for 3a, 3b, 3c, 4a, 4b, 4c, 4d would be similar}
        \If{threadIdx $\times 2 + 1 < \text{data\_len}$}{
            $dec\_by\_2\_array[\text{threadIdx}] \gets 1a + 2a + 2b$\;
        }
        \If{threadIdx $\times 3 + 2 < \text{data\_len}$}{
            $dec\_by\_3\_array[\text{threadIdx}] \gets 1a + 2a + 2b + 3a + 3b + 3c$\;
        }
        \If{threadIdx $\times 4 + 3 < \text{data\_len}$}{
            $dec\_by\_4\_array[\text{threadIdx}] \gets 1a + 2a + 2b + 3a + 3b + 3c + 4a + 4b + 4c + 4d$\;
        }
  }
\caption{Harmonic Sum Decimation Process}
\end{algorithm}

\section*{GPU Significance Calculation}

\begin{algorithm}[H]
\DontPrintSemicolon
\SetKwFunction{FMain}{power\_to\_logp}
\SetKwProg{Fn}{Device function}{:}{\KwRet}
  \Fn{\FMain{pow, dof}}{
        \uIf{dof $\geq$ pow $\times 1.05$}{
            $logp \gets 0.0$\;
        }
        \Else{
            $x \gets 1500 \times dof / pow$\;
            $f(x) \gets -4.4604059 \times 10^{-46} \times x^{16} + 9.4927863 \times 10^{-42} \times x^{15}$\;
            \Indp\Indp$- 9.1470451 \times 10^{-38} \times x^{14} + 5.2810851 \times 10^{-34} \times x^{13}$\;
            $- 2.0376166 \times 10^{-30} \times x^{12} + 5.5480334 \times 10^{-27} \times x^{11}$\;
            $- 1.0973877 \times 10^{-23} \times x^{10} + 1.5991804 \times 10^{-20} \times x^{9}$\;
            $- 1.7231488 \times 10^{-17} \times x^{8} + 1.3660070 \times 10^{-14} \times x^{7}$\;
            $- 7.8617952 \times 10^{-12} \times x^{6} + 3.2136336 \times 10^{-9} \times x^{5}$\;
            $- 9.0466418 \times 10^{-7} \times x^{4} + 0.00016945948 \times x^{3}$\;
            $- 0.021494231 \times x^{2} + 2.9515954 \times x - 755.24091$\;
            \Indm\Indm$logp \gets pow \times f(x) / 1500$\;
        }
        \KwRet $logp$\;
  }
\caption{Power to Log Probability Calculation}
\end{algorithm}

\section*{GPU Boxcar Filter}

\begin{algorithm}[H]
\DontPrintSemicolon
\SetKwProg{Fn}{Global function}{:}{\KwRet}
  \Fn{boxcar\_filter(data\_array, chunk\_size, z\_max, z\_step)}{
        \textbf{dynamic shared} $lookup\_array[\text{chunk\_size} + z\_max]$\;
        \textbf{dynamic shared} $sum\_array[\text{chunk\_size}]$\;
        \textbf{dynamic shared} $[float, int]$ $search\_reduce\_array[\text{blockDim}]$\;
        $numIterationsToCoverChunk \gets (\text{chunk\_size} + \text{blockDim} - 1) / \text{blockDim}$\;
        $numIterationsToFillLookup \gets (\text{chunk\_size} + z\_max + \text{blockDim} - 1) / \text{blockDim}$\;
        \tcp{Fill lookup\_array with data}
        \For{$i \leftarrow 0$ \KwTo $numIterationsToFillLookup$}{
            $local\_index \gets \text{threadIdx} + i \times \text{blockDim}$\;
            $global\_index \gets \text{blockIdx} \times \text{chunk\_size} + local\_index$\;
            \If{$local\_index < \text{chunk\_size} + z\_max$}{
                $lookup\_array[local\_index] \gets data\_array[global\_index]$\;
            }
        }
        \tcp{Set sum\_array to 0}
        \For{$i \leftarrow 0$ \KwTo $numIterationsToCoverChunk$}{
            $local\_index \gets \text{threadIdx} + i \times \text{blockDim}$\;
            \If{$local\_index < \text{chunk\_size}$}{
                $sum\_array[local\_index] \gets 0$\;
            }
        }
        \tcp{Search array using boxcar filters}
        \For{$z \leftarrow 0$ \KwTo $z\_max$}{
            \tcp{Apply the next boxcar filter}
            \For{$i \leftarrow 0$ \KwTo $numIterationsToCoverChunk$}{
                $local\_index \gets \text{threadIdx} + i \times \text{blockDim}$\;
                \If{$local\_index < \text{chunk\_size}$}{
                    $sum\_array[local\_index] += lookup\_array[local\_index + z]$\;
                }
            }
            \tcp{This condition on z can be replaced with a check for if the current z is a power of 2 for logarithmic sampling}
            \If{$\text{remainder}(z / z\_step) = 0$}{
                \tcp{Scan across the array for max}
                $local\_max\_power \gets 0$\;
                $local\_max\_index \gets 0$\;
                \For{$i \leftarrow 0$ \KwTo $numIterationsToCoverChunk$}{
                    $local\_index \gets \text{threadIdx} + i \times \text{blockDim}$\;
                    \If{$sum\_array[local\_index] > local\_max\_power$}{
                        $local\_max\_power \gets sum\_array[local\_index]$\;
                        $local\_max\_index \gets \text{chunk\_size} \times \text{blockIdx} + local\_index$\;
                    }
                }
                $search\_reduce\_array[\text{threadIdx}].power \gets local\_max\_power$\;
                $search\_reduce\_array[\text{threadIdx}].index \gets local\_max\_index$\;
                \tcp{Followed by block wide max reduction}
                \For{$stride \leftarrow \text{blockDim}/2, \text{blockDim}/4, \ldots, 1$}{
                    \If{$\text{threadIdx} < stride$}{
                        \If{$search\_reduce\_array[\text{threadIdx} + stride].power > search\_reduce\_array[\text{threadIdx}].power$}{
                            $search\_reduce\_array[\text{threadIdx}] \gets search\_reduce\_array[\text{threadIdx} + stride]$\;
                        }
                    }
                }
                \If{$\text{threadIdx} = 0$}{
                    \tcp{Write out local candidate to global candidate array}
                }
            }
        }
  }
\caption{Boxcar Filtering Process}
\end{algorithm}

\end{document}